\documentclass[12pt]{article}
\usepackage{graphicx}
\usepackage{color}
\usepackage[latin1]{inputenc}
\usepackage[T1]{fontenc}

\usepackage{graphicx}
\usepackage{graphics}
\usepackage{color}
\usepackage{ragged2e}
\usepackage{lipsum}
\usepackage{amsmath}
\usepackage[table]{xcolor}  
\usepackage{float}
\usepackage{cite}

\def\beq{\begin{equation}}
\def\eeq{\end{equation}}
\def\bea{\begin{eqnarray}}
\def\eea{\end{eqnarray}}
\def\nn{\nonumber}
\def\roughly#1{\mathrel{\raise.3ex\hbox
{$#1$\kern-.75em\lower1ex\hbox{$\sim$}}}}

\def\bd{B^0}

\def\pewcp{P_{EW}^{\prime C}}
\def\pewp{P'_{EW}}
\def\pewcpnp{P_{EW, NP}^{\prime C}}
\def\pewpnp{P'_{EW, NP}}

\def\btos{{\bar b} \to {\bar s}}
\def\ttoc{{\bar t} \to {\bar c}}
\def\bsll{{\bar b} \to {\bar s} \ell^+ \ell^-}
\def\bsmumu{{\bar b} \to {\bar s} \mu^+ \mu^-}
\def\bsee{{\bar b} \to {\bar s} e^+ e^-}
\def\bstautau{{\bar b} \to {\bar s} \tau^+ \tau^-}
\def\bsff{{\bar b} \to {\bar s} f {\bar f}}
\def\bsuu{{\bar b} \to {\bar s} u {\bar u}}
\def\bscc{{\bar b} \to {\bar s} c {\bar c}}
\def\bsuc{{\bar b} \to {\bar s} u {\bar c}}
\def\bscu{{\bar b} \to {\bar s} c {\bar u}}
\def\bsdd{{\bar b} \to {\bar s} d {\bar d}}
\def\bsqq{{\bar b} \to {\bar s} q {\bar q}}
\def\bsnunu{{\bar b} \to {\bar s} \nu {\bar \nu}}
\def\bctaunu{b \to c \tau^- {\bar\nu}_\tau}
\def\order{\lower 1.8ex \hbox{\LARGE\~{}}}
\def\btopik{B \to \pi K}
\def\bra#1{\left\langle #1\right|}
\def\ket#1{\left| #1\right\rangle}

\def\Npp{{\mathcal{N}^{\prime\prime}}}
\newcommand\CC{{\cal C}}

\def\ApNPCd{{\cal A}^{\prime {C}, d}}
\def\ApNPCu{{\cal A}^{\prime {C}, u}}

\def\ApNPqph{{\cal A}^{\prime,q} e^{i \Phi'_q}}
\def\ApNPCqph{{\cal A}^{\prime {C}, q} e^{i \Phi_q^{\prime C}}}
\def\ApNPCuph{{\cal A}^{\prime {C}, u} e^{i \Phi_u^{\prime C}}}
\def\ApNPCdph{{\cal A}^{\prime {C}, d} e^{i \Phi_d^{\prime C}}}
\def\ApNPuph{{\cal A}^{\prime,u} e^{i \Phi'_u}}
\def\ApNPdph{{\cal A}^{\prime,d} e^{i \Phi'_d}}
\def\ApNPcomb{{\cal A}^{\prime, comb} e^{i \Phi'}}

\begin{document}

\begin{flushright}
UdeM-GPP-TH-24-302 \\
LA-UR-24-XXXXXX
\end{flushright}

\begin{center}
\bigskip
\bigskip
{\Large \bf \boldmath Uniting Low-energy Semileptonic \\ and Hadronic Anomalies within SMEFT} \\
\bigskip
\bigskip
{\large
Alakabha Datta $^{a,}$\footnote{datta@phy.olemiss.edu},
Jacky Kumar $^{b,}$\footnote{ jacky.kumar@lanl.gov},
Suman Kumbhakar $^{c,}$\footnote{suman.kumbhakar@umontreal.ca} \\
and David London $^{c,}$\footnote{london@lps.umontreal.ca}
}
\end{center}

\begin{flushleft}
~~~~~~~~~~~$a$: {\it Department of Physics and Astronomy,}\\
~~~~~~~~~~~~~~~{\it 108 Lewis Hall, University of Mississippi, Oxford, MS 38677, USA}\\
~~~~~~~~~~~$b$: {\it Theoretical Division, Los Alamos National Laboratory,}\\
~~~~~~~~~~~~~~~{\it Los Alamos, NM 87545, USA}\\
~~~~~~~~~~~$c$: {\it Physique des Particules, Universit\'e de Montr\'eal,}\\
~~~~~~~~~~~~~~~{\it 1375, ave Th\'er\`ese-Lavoie-Roux, Montr\'eal, QC, Canada H2V 0B3}\\
\end{flushleft}

\begin{center}
\bigskip (\today)
\vskip0.5cm {\Large Abstract\\} \vskip3truemm
\parbox[t]{\textwidth}{Two categories of four-fermion SMEFT operators are semileptonic (two quarks and two leptons) and hadronic (four quarks). At tree level, an operator of a given category contributes only to processes of the same category. However, when the SMEFT Hamiltonian is evolved down from the new-physics scale to low energies using the renormalization-group equations (RGEs), due to operator mixing this same SMEFT operator can generate operators of the other category at one loop. Thus, to search for a SMEFT explanation of a low-energy anomaly, or combination of anomalies, one must: (i) identify the candidate semileptonic and hadronic SMEFT operators, (ii) run them down to low energy with the RGEs, (iii) generate the required low-energy operators with the correct Wilson coefficients, and (iv) check that all other constraints are satisfied. In this paper, we illustrate this method by finding all SMEFT operators that, by themselves, provide a combined explanation of the (semileptonic) $\bsll$ anomalies and the (hadronic) $\btopik$ puzzle.}

\end{center}


\thispagestyle{empty}
\newpage
\setcounter{page}{1}
\baselineskip=14pt

\section{Introduction}

The standard model (SM) of particle physics has been enormously
successful in describing the physics up to energy scales of $O({\rm
  TeV})$. It has made many predictions, almost all of which have been
verified. Even so, it is not complete, as it cannot account for
several observations (neutrino masses, dark matter, the baryon
asymmetry of the universe, etc.). There must be physics beyond the SM.

No new particles have been observed at the LHC, so we are forced to
conclude that this new physics (NP), whatever it is, must be
heavy. The modern, model-independent approach to analyzing such NP
uses effective field theories (EFTs). When the NP is integrated out,
one obtains the SMEFT \cite{Buchmuller:1985jz, Grzadkowski:2010es, Brivio:2017vri}, an
EFT that contains only the SM particles and obeys the SM gauge symmetry,
$SU(3)_C \times SU(2)_L \times U(1)_Y$. The leading-order
(dimension-4) terms are those of the SM; higher-order terms are
suppressed by powers of the NP scale $\Lambda$.

These higher-order, non-SM terms can provide new contributions to
low-energy processes. Their presence leads to an indirect signal of NP
when the measurement of an observable in a given process disagrees
with the prediction of the SM. Indeed, whenever such an anomaly is
observed, one wants to identify the type of NP that could lead to this
effect. One method is to build models. But the model-independent
approach is to determine the SMEFT operator(s) that can modify the
low-energy process. This is a complicated procedure because of
operator mixing. A SMEFT operator is defined at the scale
$\Lambda$. When the Hamiltonian is evolved down to low energies using
the renormalization-group equations (RGEs), many operators are
generated at one loop. This means that there are a number of SMEFT
operators that could affect the process in question. But it also means
that, if a given SMEFT operator creates a deviation in one low-energy
process, it is likely to also lead to a discrepancy in another process
(or more). Using the RGEs, we can determine which low-energy processes 
can be affected by each SMEFT operator (see, for example, Refs.~\cite{Buras:2014fpa, Kumar:2021yod, Aebischer:2020dsw, Aebischer:2020lsx, Aebischer:2020mkv, Bobeth:2017ecx, Bobeth:2017xry, Alok:2021ydy, Cirigliano:2023nol}).

At present, there are a number of anomalies in $B$ decays. Some are in
semileptonic decays \cite{London:2021lfn}, while others are in
hadronic decays \cite{Beaudry:2017gtw, Bhattacharya:2021shk,
  Bhattacharya:2022akr, Amhis:2022hpm, Biswas:2023pyw}. Various NP
solutions have been proposed as explanations of the individual
anomalies, and as simultaneous explanations of all the semileptonic
anomalies \cite{Bhattacharya:2014wla, Greljo:2015mma, Calibbi:2015kma,
  Barbieri:2015yvd, Boucenna:2016qad, Bhattacharya:2016mcc,
  Crivellin:2017zlb, Buttazzo:2017ixm, Kumar:2018kmr,
  Angelescu:2021lln}. However, nobody has looked for a combined
solution to one of each type of anomaly. This involves finding NP that
contributes to both semileptonic and hadronic $\btos$ transitions. In
this paper, we show that this can be done within SMEFT: when one takes
a single SMEFT operator and runs it down to the scale $m_b$, both
semileptonic and hadronic $\btos$ operators can be generated. It is
then necessary to check that the Wilson coefficients of these
low-energy operators take the right values to explain the semileptonic
and hadronic $B$ anomalies. Other operators will also be generated, so
that constraints from other processes must be taken into account. This
procedure is quite general -- it can be applied to any low-energy
processes that exhibit a discrepancy with the SM (not just $B$ decays).

In order to illustrate how this method works, here we focus on two
specific $B$ anomalies. First, for more than ten years, there have
been a number of measurements of observables involving the
semileptonic decay $\bsll$ ($\ell = \mu, e$) that are in disagreement
with the predictions of the SM.  Until 2021, these could be explained
if only $\bsmumu$ receives NP contributions
\cite{London:2021lfn}. However, in late 2022, LHCb announced that it
had remeasured the ratios $R_K$ and $R_{K^*}$, which test
lepton-flavour universality, and found that they agree with the SM
\cite{LHCb:2022qnv, LHCb:2022vje}. Now the most promising explanation
is that the NP contributes equally to $\bsmumu$ and $\bsee$
\cite{Greljo:2022jac, Alguero:2023jeh, Hurth:2023jwr}.  (A model to
generate equal contribution to $\bsmumu$ and $\bsee$ via four quark
operators was proposed much earlier \cite{Datta:2013kja}.)

The second $B$-decay anomaly has been around even longer, about 20
years: it is the $\btopik$ puzzle (see Refs.~\cite{Beaudry:2017gtw,
  Bhattacharya:2021shk} and references therein). Here the amplitudes
for the four decays $B^+ \to \pi^+ K^0$, $B^+ \to \pi^0 K^+$, $B^0 \to
\pi^- K^+$ and $B^0 \to \pi^0 K^0$ obey a quadrilateral isospin
relation.  However, the measurements of the observables in these
decays are not completely consistent with one another -- there is a
discrepancy at the level of $\sim 3\sigma$.

Because $\bsll$ and $\btopik$ decays both involve $\btos$ transitions,
it is natural to look for a simultaneous explanation of the two
anomalies. In this paper, we search for a single SMEFT operator that,
when run down from the scale $\Lambda$ to $m_b$, generates both
$\bsll$ and $\bar b \to \bar s q \bar q$ ($q=u,d$) operators with
Wilson coefficients of the right values to account for the $\bsll$ and
$\btopik$ anomalies. Constraints from other observables, such as
$\bsnunu$, $\Delta M_s$, etc., must also be satisfied. As we will see,
there are a handful of four-quark SMEFT operators that can do this.

We begin in Sec.~2 with a review of the $\bsll$ anomalies and the
$\btopik$ puzzle. In Sec.~3, we examine the $\bsll$ anomalies in the
context of the SMEFT. We find that one semileptonic and six four-quark
SMEFT operators can account for the $\bsll$ results, while satisfying
all other constraints. The $\btopik$ puzzle is studied in the context
of SMEFT in Sec.~4. Of the seven SMEFT operators identified in Sec.~3,
we find that three four-quark operators can also give a good fit to
the $\btopik$ data, while one four-quark operator provides a passable
fit.  We conclude in Sec.~5.

\section{\boldmath $B$-Decay Anomalies}

\subsection{\boldmath $\bsll$}

Over the years, there have been many analyses in which a global fit to
all the $\bsll$ data was performed with the aim of determining which
NP scenarios are preferred to explain the anomalies. The method is as
follows.

At the scale $m_b$, the physics is described by the WET (weak
effective theory), obtained by integrating out the heavy degrees of
freedom. These include all SM and NP particles heavier than the $b$
quark. The WET operators obey the $SU(3)_C \times U(1)_{em}$ gauge
symmetry. The effective Lagrangian describing $\bsll$ transitions is
given by \cite{Buchalla:1995vs}
\beq
{\cal L}_{\rm eff}= \frac{4G_F}{\sqrt{2}} \, V_{tb}V_{ts}^* \, \frac{e^2}{16 \pi^2} \sum_i {\cal C}_{i}  {\cal O}_i ~,
\label{heff}
\eeq
where $V_{ij}$ are elements of the Cabibbo-Kobayashi-Maskawa (CKM)
matrix and the WET operators are
\bea 
& {\mathcal{O}}_{9\ell}^{(\prime)} = (\bar{s} \gamma_{\mu} P_{L(R)} b) (\bar{\ell} \gamma^\mu \ell) ~~,~~~~
{\mathcal{O}}_{10\ell}^{(\prime)} = (\bar{s} \gamma_{\mu} P_{L(R)} b) (\bar{\ell} \gamma^\mu \gamma_5 \ell) ~, & \nn\cr
& {\mathcal{O}}_7^{(\prime)} = \frac{m_b}{e} \, (\bar{s} \sigma_{\mu \nu} P_{R(L)} b) F^{\mu \nu} ~, & \nn\cr
& {\mathcal{O}}_{S\ell}^{(\prime)} = (\bar{s} P_{R(L)} b)(\bar{\ell} \ell) ~~,~~~~
{\mathcal{O}}_{P\ell}^{(\prime)} = (\bar{s} P_{R(L)} b)(\bar{\ell} \gamma_5 \ell) ~, &
\label{eq:bsll-wet}
\eea
with $P_{L(R)} = \frac12 (1 -(+) \gamma_5)$.
In the above, only operators that are generated at dimension 6 in the
SMEFT have been kept. This
requirement also imposes the following conditions on the Wilson
coefficients (WCs): $C_{S\ell} = -C_{P\ell}$ and $C'_{S\ell} =
C'_{P\ell}$.

${\cal H}_{\rm eff}$ is valid at the scale $m_b$. All information
about the heavy particles that have been integrated out is encoded in
the WCs of the operators, ${\cal C}_{i}$. Of the eight operators
describing $\bsll$ transitions, the SM contributes mainly to
${\mathcal{O}}_7$, ${\mathcal{O}}_{9\ell}$ and
${\mathcal{O}}_{10\ell}$, but the NP can contribute to all of them.

There are a great many $\bsll$ observables. Some involve only the
$\bsmumu$ transition. These include branching ratios (e.g., ${\cal
  B}({B^+ \to K^+\mu^+\mu^-})$, ${\cal B}(B_s \to \mu^+ \mu^-)$, etc.)
and angular observables in four-body decays (e.g., $B \to K^*(\to
K\pi)\mu^+\mu^-$, $B_s \to \phi(\to K^+K^-) \mu^+\mu^-$). Others
measure lepton-flavour-universality violation: 
\beq
    R_K = \frac{{\cal B}(B^+ \to K^+ \mu^+ \mu^-)}{{\cal B}(B^+ \to K^+ e^+ e^-)} ~~,~~~~ 
    R_{K^*} = \frac{{\cal B}(B \to K^{*} \mu^+ \mu^-)}{{\cal B}(B \to K^{*} e^+ e^-)} ~~,~~~~ 
    R_{\phi} = \frac{{\cal B}(B_s \to \phi \, \mu^+ \mu^-)}{{\cal B}(B_s \to \phi \, e^+ e^-)} ~.
\eeq
The SM predicts that all of these ratios equal 1 (to a very good
approximation).

All observables can be written as a function of the WCs in
Eq.~(\ref{heff}).  These WCs are separated into their SM and NP
contributions: $\CC_i = \CC_i^{\rm SM} + \CC_i^{\rm NP}$. By
performing fits to the data, using subsets of the $\CC_i^{\rm NP}$ as
free parameters, it is possible to determine which (combinations of)
$\CC_i^{\rm NP}$ best explain the data, and what their best-fit values
are.

Following the LHCb announcement in 2022 of the new results for $R_K$
and $R_{K^*}$ that agreed with the SM, three global-fit analyses
appeared, Refs.~\cite{Greljo:2022jac, Alguero:2023jeh, Hurth:2023jwr}.
All found that the NP scenario that best fits the data involves a
single WC, $\CC_{9\mu}^{\rm NP} = \CC_{9e}^{\rm NP} \equiv C_9^{\rm
  U}$. And they found similar best-fit values: $C_9^{\rm U} = -0.77
\pm 0.21$ \cite{Greljo:2022jac}, $-1.17^{+0.16}_{-0.17}$
\cite{Alguero:2023jeh}, $-1.18 \pm 0.19$ \cite{Hurth:2023jwr}. In our
analysis, we require that $C_9^{\rm U}$ equal the last of these
best-fit values.

\subsection{\boldmath $\btopik$}
\label{SecpiK}

Here we follow and update the analysis of Ref.~\cite{Beaudry:2017gtw}.
There are four $\btopik$ decays: $B^+ \to \pi^+ K^0$ (designated as
$+0$ below), $B^+ \to \pi^0 K^+$ ($0+$), $\bd \to \pi^- K^+$ ($-+$)
and $\bd \to \pi^0 K^0$ ($00$). Their amplitudes obey a quadrilateral
isospin relation:
\beq
\sqrt{2} A^{00} + A^{-+} = \sqrt{2} A^{0+} + A^{+0} ~.
\eeq
Using these decays, nine observables have been measured: the four
branching ratios, the four direct CP asymmetries $A_{CP}$, and the
mixing-induced indirect CP asymmetry $S_{CP}$ in $\bd\to \pi^0K^0$.
The latest data are shown in Table \ref{tab:data}.

\begin{table}[h!]
    \centering
    \begin{tabular}{|c|c|c|c|}
    \hline
     Decay   & $\mathcal{B}\times 10^{-6}$& $A_{CP}$ & $S_{CP}$ \\ 
    \hline
    $B^+\to \pi^+K^0$ & $23.9\pm 0.6$ & $-0.003\pm 0.015$ & \\
    \hline
    $B^+\to \pi^0K^+$ & $13.2\pm 0.4$ & $0.027\pm 0.012$ & \\
    \hline
    $B^0\to \pi^-K^+$ & $20.0\pm 0.4$ & $-0.0831\pm 0.0031$ & \\
    \hline
    $B^0\to \pi^0K^0$ & $10.1\pm 0.4$ & $0.00\pm 0.08$ & $0.64\pm 0.13$ \\
    \hline
    \end{tabular}
\caption{Branching ratios, direct CP asymmetries $A_{CP}$, and
  mixing-induced CP asymmetry $S_{CP}$ (if applicable) for the four
  $\btopik$ decay modes. The data are taken from the Particle Data
  Group 2024 \cite{pdg}.}
\label{tab:data}
\end{table}

Within the diagrammatic approach of Refs.~\cite{Gronau:1994rj,
  Gronau:1995hn}, $B$-decay amplitudes are expressed in terms of six
diagrams\footnote{The annihilation, exchange and penguin-annihilation
  diagrams are neglected, as they are expected to be very small in the
  SM.}: the color-favored and color-suppressed tree amplitudes $T'$
and $C'$, the gluonic penguin amplitudes $P'_{tc}$ and $P'_{uc}$, and
the color-favored and color-suppressed electroweak penguin amplitudes
$\pewp$ and $\pewcp$ (the primes on the amplitudes indicate $\btos$
transitions). The $\btopik$ amplitudes are given by
\bea
\label{fulldiagrams}
A^{+0} &=& -P'_{tc} + P'_{uc} e^{i\gamma} -\frac13
\pewcp ~, \nn\\
\sqrt{2} A^{0+} &=& -T' e^{i\gamma} -C' e^{i\gamma}
+P'_{tc} -~P'_{uc} e^{i\gamma} -~\pewp -\frac23 \pewcp ~,
\nn\\
A^{-+} &=& -T' e^{i\gamma} + P'_{tc} -P'_{uc}
e^{i\gamma} -\frac23 \pewcp ~, \nn\\
\sqrt{2} A^{00} &=& -C' e^{i\gamma} - P'_{tc} +P'_{uc}
e^{i\gamma} - \pewp -\frac13 \pewcp ~.
\eea
The weak-phase dependence (including the minus sign from $V_{tb}^*
V_{ts}$ [in $P'_{tc}$]) is written explicitly, so that the diagrams
contain both strong phases and the magnitudes of the CKM matrix
elements. The amplitudes for the CP-conjugate processes are obtained
from the above by changing the sign of the weak phase $\gamma$.

For the fit to the data, we use the following conditions regarding the
unknown theoretical parameters (we refer the reader to
Ref.~\cite{Beaudry:2017gtw} for explanations). (i) The diagrams
$\pewp$ and $\pewcp$ are not independent -- to a good approximation,
they can be related to $T'$ and $C'$ within the SM using flavor
$SU(3)$ symmetry \cite{Neubert:1998pt, Neubert:1998jq, Gronau:1998fn}.
(ii) We fix $|C'/T'| = 0.2$, its preferred theoretical value
\cite{Beneke:2001ev, Bell:2007tv, Bell:2009nk, Beneke:2009ek,
  Bell:2015koa}.  (iii) We neglect $P'_{uc}$ \cite{Kim:2007kx}. (iv)
We adopt the convention that the strong phase $\delta_{T'} = 0$
\cite{Datta:2004jm}. (v) For the weak phases, we include the
constraints from direct measurements: $\beta = (22.2\pm 0.7)^{\circ}$,
$\gamma = (66.2^{+3.4}_{-3.6})^{\circ}$ \cite{HFLAV:2022esi}.

With the above conditions, there are four independent theoretical
parameters in the amplitudes: $|T'|$, $|P'_{tc}|$, and the two strong
phases $\delta_{P'_{tc}}$ and $\delta_{C'}$. With nine $\btopik$
observables, a fit can be done; the results are given in Table
\ref{tab:SMfitCT0.2}. The fit is poor: $\chi^2_{\rm min}/{\rm d.o.f.}
= 16.9/5$, corresponding to a p-value of 0.5\%. This is the $\btopik$
puzzle: there is a discrepancy with the SM at the level of $2.8\sigma$.

\begin{table}[h!]
    \centering
    \begin{tabular}{|c|c|}
\hline
\multicolumn{2}{|c|}{$\chi^2_{\rm min}/{\rm d.o.f.} = 16.9/5$,}\\
\multicolumn{2}{|c|}{p-value $= 0.005$}\\
 \hline
Parameter & Best-fit value \\ 
\hline
$\gamma$ & $(64.74 \pm 2.86)^\circ$ \\
    \hline
$\beta$ & $(22.12 \pm 0.69)^\circ$ \\
    \hline
    $|T^{\prime}|$ & $8.4\pm 0.9$\\
    \hline
    $|P^{\prime}_{tc}|$ & $51.2\pm 0.4$\\
    \hline
    $\delta_{P^{\prime}_{tc}}$ & $(-13.2 \pm 1.7)^\circ$ \\
    \hline
    $\delta_{C^{\prime}}$ & $(260.7 \pm 16.0)^\circ$\\
\hline
\end{tabular}
\caption{$\chi^2_{\rm min}/{\rm d.o.f.}$ and best-fit values of unknown
  parameters in amplitudes of Eq.~(\ref{fulldiagrams}). Constraints:
  $\btopik$ data, measurements of $\beta$ and $\gamma$, theoretical
  inputs $|C'/T'| = 0.2$, $P'_{uc} = 0$.}
\label{tab:SMfitCT0.2}
\end{table}

We now turn to NP. At the level of the WET effective Hamiltonian, the
NP operators that contribute to the $\btopik$ amplitudes take the form
${\cal O}_{NP}^{ij,q} \sim {\bar s} \Gamma_i b \, {\bar q} \Gamma_j q$
($q = u,d$), where $\Gamma_{i,j}$ represent Lorentz structures, and
color indices are suppressed. The NP contributions to $\btopik$ are
encoded in the matrix elements $\bra{\pi K} {\cal O}_{NP}^{ij,q}
\ket{B}$. In general, each matrix element has its own NP weak and
strong phases.

In Ref.~\cite{Beaudry:2017gtw}, it was argued that the NP strong
phases are negligible (see also Ref.~\cite{Datta:2004jm}), so that one
can combine many NP matrix elements into a single NP amplitude, with a
single weak phase. There are two classes of such NP amplitudes,
differing only in their color structure:
\bea \label{eq: amplitudes}
\sum \bra{\pi K} {\bar s}_\alpha \Gamma_i b_\alpha \, {\bar q}_\beta \Gamma_j q_\beta \ket{B} &\equiv& \ApNPqph
~, \nn\\
\sum \bra{\pi K} {\bar s}_\alpha \Gamma_i b_\beta \, {\bar q}_\beta \Gamma_j q_\alpha \ket{B} &\equiv& \ApNPCqph
~~,~~ q = u,d ~.
\eea
Although there are four NP matrix elements that contribute to
$\btopik$ decays, only three combinations appear in the amplitudes:
$\ApNPcomb \equiv - \ApNPuph + \ApNPdph$, $\ApNPCuph$, and
$\ApNPCdph$.  The $\btopik$ amplitudes can now be written in terms of
the SM diagrams and these NP matrix elements:
\bea
\label{BpiKNPamps}
A^{+0} &=& -P'_{tc} -\frac13 \pewcp + \ApNPCdph ~, \nn\\
\sqrt{2} A^{0+} &=& P'_{tc} - T' \, e^{i\gamma} - \pewp -C' \, e^{i\gamma} -\frac23 \pewcp + \ApNPcomb - \ApNPCuph ~, \nn\\
A^{-+} &=& P'_{tc} - T' \, e^{i\gamma} -\frac23 \pewcp - \ApNPCuph ~, \nn\\
\sqrt{2} A^{00} &=& -P'_{tc} - \pewp -C' \, e^{i\gamma} -\frac13 \pewcp + \ApNPcomb + \ApNPCdph ~.
\eea

We note in passing that a better understanding of these NP
contributions can be found with a change of basis \cite{Baek:2009pa}:
\bea
\pewpnp \, e^{i \Phi'_{EW}} & \equiv & \ApNPuph - \ApNPdph ~, \nn\\
P'_{NP} \, e^{i \Phi'_{P}} & \equiv & \frac13 \ApNPCuph\ + \frac23 \ApNPCdph~, \nn\\
\pewcpnp \, e^{i \Phi^{\prime {C}}_{EW}} & \equiv & \ApNPCuph\ - \ApNPCdph ~.
\label{NPPoperators}
\eea
In order, these three terms correspond to the inclusion of NP in the
color-allowed electroweak penguin, the gluonic penguin, and the
color-suppressed electroweak penguin amplitudes.

In the most general case, each of the three independent NP matrix
elements has its own weak phase.  There are therefore 10 parameters in
the $\btopik$ amplitudes: 5 magnitudes of diagrams, 2 strong phases,
and 3 NP weak phases. This is greater than the number of observables
(9), so a fit cannot be performed.

However, suppose that these NP matrix elements are the low-energy
remnants of the same SMEFT operator. That is, we begin at the
high-energy NP scale with a single SMEFT operator and evolve it down
to the scale $m_b$ with the RGEs. This will
generate a variety of $\bsqq$ operators ($q=u,d$), and these can be
used to compute the NP matrix elements (magnitudes and weak phases).
A fit can then be performed to see if the addition of these NP
contributions removes the discrepancy with the SM. As we will see in
the next section, this is indeed possible. Not only that, but this
same SMEFT operator will generate ${\mathcal{O}}_{9\ell}$ with
$C_9^{\rm U} = -1.18 \pm 0.19$!

\section{\boldmath SMEFT and $\bsll$}

In the previous section, we saw that the most promising NP scenario to
explain the $\bsll$ anomalies involves a new contribution to the WET
operator ${\mathcal{O}}_{9\ell} = (\bar{s} \gamma_{\mu} P_L b)
(\bar{\ell} \gamma^\mu \ell)$ with a value of the WC $C_9^{\rm U} =
-1.18 \pm 0.19$. We also saw that any solution to the $\btopik$ puzzle
must involve NP contributions to WET $\bsqq$ operators ($q=u,d$). And
we intimated that combined explanations can be found in which the WET
operators all arise from the same SMEFT operator.

In order to see how this comes about, we need to examine (i)
properties of the SMEFT, (ii) the SMEFT solution of the $\bsll$
anomalies, and (iii) how this $\bsll$ SMEFT solution connects to the
$\btopik$ puzzle. In this section, we focus on the first two points;
the connection to $\btopik$ will be discussed in Sec.~\ref{SimExp}.

\subsection{SMEFT}

Up to dimension 6, the SMEFT Lagrangian can be written as 
\beq
\label{eq:1}
\mathcal{L}_{\rm SMEFT} =
\mathcal{L}_{\textrm{SM}}+
\sum_{ { Q}_a^\dagger = {Q}_a} {C}_a {Q}_a 
+ \sum_{{Q}_a^\dagger \ne {Q}_a} \left ( {C}_a {Q}_a + {C}_a^* {Q}_a^\dagger  \right   ).
\eeq 
Here $\mathcal{L}_{\textrm{SM}}$ is the dimension-4 part of the SMEFT
Lagrangian. The $Q_a$ are dimension-6 operators; the ${C}_a$ are their
Wilson coefficients. All the $Q_a$ were derived in
Refs.~\cite{Grzadkowski:2010es}; this set of
operators is known as the Warsaw basis.

The SMEFT Lagrangian describes the physics from the scale $\Lambda$
down to the weak scale. At the weak scale, the $SU(3)_C \times SU(2)_L
\times U(1)_Y$ gauge symmetry is broken to $SU(3)_C \times U(1)_{em}$
and the WET Lagrangian is now applicable. One can work out which SMEFT
operators generate a given WET operator. These are known as matching
conditions.

But note that SMEFT operators are defined in the flavour (gauge)
basis, whereas the WET operators use the physical mass basis. Thus,
matching SMEFT and WET operators requires us to take into account the
unitary transformations from the flavour to the mass basis. We adopt
the convention that the flavour and mass eigenstates are the same for
the charged leptons and down-type quarks. For the neutrinos, the
difference between the two bases is unimportant, since they are not
detected. However, for up-type quarks,
\beq
U_{Li} = V^\dagger_{ij} \, U^0_{Lj} ~,
\label{uquarkdef}
\eeq
where $U = (u, c, t)^T$, $V$ is the CKM matrix, and the superscript 0
(lack of superscript) indicates the flavour (mass) eigenstates.

Finally, both the SMEFT and WET are energy-dependent. That is,
although the SMEFT is applicable from scale $\Lambda$ down to the weak
scale, the coefficients of the operators are not constant. Their
values change due to operator mixing as the SMEFT evolves from one
energy scale to another. This can all be computed using the RGEs.
This also applies to the WET.
RGEs can be used to calculate how the WET coefficients change as we
evolve from the weak scale down to $m_b$.

\subsection{\boldmath $\bsll$ in SMEFT}

As discussed above, the preferred NP solution to the $\bsll$ anomalies
is that there is a new contribution to the WET operator
${\mathcal{O}}_{9\ell} = (\bar{s} \gamma_{\mu} P_L b) (\bar{\ell}
\gamma^\mu \ell)$ with a value of $\CC_{9\mu}^{\rm NP} = \CC_{9e}^{\rm
  NP} \equiv C_9^{\rm U} = -1.18 \pm 0.19$.  The question
is: what is the origin of this lepton-flavour-universal (LFU) effect?
(Note that here ``universal'' refers to only $e$ and $\mu$.)

\subsubsection{WET-SMEFT matching conditions}

There are five vector SMEFT operators that involve two quarks and two leptons:
\bea
& Q^{(1)}_{\ell q} = (\bar{\ell}_i \gamma_{\mu}\ell_j)(\bar{q}_k \gamma^{\mu}q_{\ell}) ~~,~~~~
Q^{(3)}_{\ell q} = (\bar{\ell}_i \gamma_{\mu}\tau^{I}\ell_j)(\bar{q}_k \gamma^{\mu}\tau^{I}q_{\ell}) ~, & \nn\\
& Q_{\ell d} = (\bar{\ell}_i \gamma_{\mu}\ell_j)(\bar{d}_k \gamma^{\mu}d_{\ell}) ~~,~~~~ 
Q_{qe} = (\bar{q}_i \gamma_{\mu}q_j) (\bar{e}_k \gamma^{\mu}e_{\ell}) ~, & \nn\\
& Q_{e d} = (\bar{e}_i \gamma_{\mu}e_j)(\bar{d}_k \gamma^{\mu}d_{\ell}) ~. &
\label{2q2lvecSMEFT}
\eea
Here $q$ and $\ell$ are LH $SU(2)_L$ doublets, while $d$ and $e$ are
RH $SU(2)_L$ singlets. $i,j,k,l$ are flavour indices -- they indicate
the generation of the fermion field. There are three vector SMEFT operators involving the Higgs field:
\bea
& Q_{\varphi q}^{(1)} = (\varphi^\dagger i \overleftrightarrow{D}_\mu \varphi ) ({\bar q}_i \gamma^\mu q_j) ~~,~~~~
Q_{\varphi q}^{(3)} = (\varphi^\dagger i \overleftrightarrow{D}_\mu^I \varphi ) ({\bar q}_i \tau^I \gamma^\mu q_j) ~, & \nn\\
& Q_{\varphi d} = (\varphi^\dagger i \overleftrightarrow{D}_\mu \varphi ) ({\bar d}_i \gamma^\mu d_j) ~. &
\eea

There are four vector $\bsll$ WET operators:
${\mathcal{O}}_{9\ell}^{(\prime)}$ and
${\mathcal{O}}_{10\ell}^{(\prime)}$ [see Eq.~(\ref{eq:bsll-wet})].
The matching conditions between the WET and SMEFT operators are given by
\bea
C_{9,\ell} &=& \frac{1}{2\mathcal{N}} \left( [C_{\ell q}^{(1)}]_{ll23} + [{C}_{\ell q}^{(3)}]_{ll23} + [C_{q e}]_{23ll} + 
\left[ {C}_{\varphi q}^{(1)23} + {C}_{\varphi q}^{(3)23} \right] (-1 + 4 \sin^2 \theta_W ) \right)
~, \nn \\
C_{10,\ell} &=& \frac{1}{2\mathcal{N}} 
\left( [C_{q e}]_{23ll} - [C_{\ell q}^{(1)}]_{ll23} - [C_{\ell q}^{(3)}]_{ll23} + 
\left[ {C}_{\varphi q}^{(1)23} + {C}_{\varphi q}^{(3)23} \right] \right) ~, \nn\\
C'_{9,\ell} &=& \frac{1}{2\mathcal{N}} 
\left( [C_{\ell d}]_{ll23} + [C_{e d}]_{ll23} + 
{C}_{\varphi d}^{23} (-1 + 4 \sin^2 \theta_W ) \right) ~, \nn\\
C'_{10,\ell} &=& \frac{1}{2\mathcal{N}} 
\left( [C_{e d}]_{ll23} - [{C}_{\ell d}]_{ll23} + 
{C}_{\varphi d}^{23} \right) ~,
\label{matching}
\eea
where $\mathcal{N} = {4G_F \over \sqrt{2}} V_{tb} V_{ts}^* {e^2 \over
  {16\pi^2}}$.  For $\ell = e$, $\mu$ and $\tau$, the indices $ll$ are
respectively $11$, $22$ and $33$. Note that, since $\sin^2 \theta_W = 0.223$, the factor $-1 + 4 \sin^2 \theta_W = -0.11$, so that the contribution of ${C}_{\varphi q}^{(1)23} + {C}_{\varphi q}^{(3)23}$ to $C_{9,\ell}$ is suppressed compared to its contribution to $C_{10,\ell}$.

These matching conditions are very informative. First, for a given
$\ell$ ($e$ or $\mu$), in order to have only $C_{9,\ell}$ nonzero, we
require particular relations among $[C_{e d}]_{ll23}$, $[{C}_{\ell d}]_{ll23}$ and 
$C_{\varphi d}^{23}$ (for $C'_{9,\ell}$ and $C'_{10,\ell}$), and among
$[C_{q e}]_{23ll}$, $[C_{\ell q}^{(1)}]_{ll23}$, $[C_{\ell q}^{(3)}]_{ll23}$,
$C_{\varphi q}^{(1)23}$ and $C_{\varphi q}^{(3)23}$ (for $C_{10,\ell}$).
Second, in order to have $C_{9,\mu}=C_{9,e}$, an additional equality among WCs is needed, namrly
$[C_{\ell q}^{(1)}]_{1123} + [{C}_{\ell q}^{(3)}]_{1123} + [C_{q e}]_{2311} 
= [C_{\ell q}^{(1)}]_{2223} + [{C}_{\ell q}^{(3)}]_{2223} + [C_{q e}]_{2322}$. 
In other words, we
require  many special relations among the WCs of operators that are
{\it a-priori} independent.  While this is logically possible, it is
somewhat ``fine-tuned.''

\subsubsection{One-loop RGE running}

Fortunately, there is an alternative, more compelling explanation.
Suppose we have a single SMEFT operator that contributes to $\bsff$,
where $f$ is a light SM fermion. As the Hamiltonian is evolved down to
low energies, renormalization-group running naturally generates
$\bsll$ transitions at one loop via the exchange of an off-shell
neutral gauge boson ($GB$) \cite{Bobeth:2011st}: $f {\bar f} \to GB^*
\to \ell^+ \ell^-$.  For the running from $\Lambda$ to the weak scale,
$GB = W_3^0, B^0$, while for the running from the weak scale to $m_b$,
$GB$ is dominantly a photon. This produces the operator
${\mathcal{O}}_{9\ell}$.  And since the $\mu$ and $e$ have the same
quantum numbers ($I_3$, $Y$, $Q_{em}$), we automatically have
$\CC_{9\mu}^{\rm NP} = \CC_{9e}^{\rm NP} \equiv C_9^{\rm U}$. This can
then potentially account for the $\bsll$ anomalies.

SMEFT operators that contribute to $\bsff$ come in two categories: (i)
semileptonic operators (two quarks, two leptons), and (ii) four-quark
operators. These can have scalar or vector Lorentz
structures. However, we also require that the renormalization-group
running of an operator generate an LFU ${\mathcal{O}}_{9\ell}$ at the
scale $m_b$. This excludes the SMEFT operators with a scalar Lorentz
structure.

The list of vector semileptonic SMEFT operators that can potentially
generate an LFU ${\mathcal{O}}_{9\ell}$ is given in
Table~\ref{tab:semilepSMEFT} [see also
  Eq.~(\ref{2q2lvecSMEFT})]. Although the flavour indices are
suppressed, we can easily deduce the possibilities. Because we have a
$b\to s$ transition, the quark current must have indices 2 and 3. And
because we want $\CC_{9\mu}^{\rm NP} = \CC_{9e}^{\rm NP}$, the indices
for the leptonic current can only be 3 and 3. There are a total of 5
possible semileptonic operators.

\begin{table}[H] 
\begin{center}
\begin{tabular}{|c|c|c|c|}
   \hline
 operator  & definition & chirality& flavour structure \\
   \hline
   $Q^{(1)}_{\ell q}$  & ($\bar{\ell}_i \gamma_{\mu}\ell_j$)($\bar{q}_k \gamma^{\mu}q_{\ell}$) & ($\bar{L}L$)($\bar{L}L$)& $3323$ \\ 
   $Q^{(3)}_{\ell q}$  & ($\bar{\ell}_i \gamma_{\mu}\tau^{I}\ell_j$)($\bar{q}_k \gamma^{\mu}\tau^{I}q_{\ell}$) & & $3323$\\ 
   \hline
   $Q_{\ell d}$  & ($\bar{\ell}_i \gamma_{\mu}\ell_j$)($\bar{d}_k \gamma^{\mu}d_{\ell}$) & ($\bar{L}L$)($\bar{R}R$) & $3323$ \\  
   $Q_{qe}$  & ($\bar{q}_i \gamma_{\mu}q_j$) ($\bar{e}_k \gamma^{\mu}e_{\ell}$)& & $2333$\\ 
   \hline
   $Q_{e d}$  & ($\bar{e}_i \gamma_{\mu}e_j$)($\bar{d}_k \gamma^{\mu}d_{\ell}$) & ($\bar{R}R$)($\bar{R}R$) & $3323$\\
   \hline
\end{tabular}
\caption{\small The list of semileptonic SMEFT operators that can
  potentially generate an LFU ${\mathcal{O}}_{9\ell}$ at scale $m_b$.}
\label{tab:semilepSMEFT}
\end{center}
\end{table}

Table 4 contains the list of possible vector four-quark SMEFT
operators. One of the quark currents must have flavour indices 2 and
3, while the other quark current has indices $i$ and $i$, $i =
1,2,3$. For the $Q^{(1)}_{q d}$ and $Q^{(8)}_{qd}$ operators, this
flavour assignment can be done in two different ways. There are
therefore a total of 33 possible four-quark operators.

\begin{table}[H]
\begin{center}
\begin{tabular}{|c|c|c|c|}
   \hline 
  operator & definition & chirality & flavour structure\\
   \hline
   $Q^{(1)}_{q q}$  & ($\bar{q}_i \gamma_{\mu}q_j$)($\bar{q}_k \gamma^{\mu}q_{\ell}$) & ($\bar{L}L$)($\bar{L}L$)& $ii23$ \\ 
   $Q^{(3)}_{q q}$  & ($\bar{q}_i \gamma_{\mu}\tau^{I}q_j$)($\bar{q}_k \gamma^{\mu}\tau^{I}q_{\ell}$) & & $ii23$\\ 
   \hline

   $Q^{(1)}_{q d}$  & ($\bar{q}_i \gamma_{\mu}q_j$)($\bar{d}_k \gamma^{\mu}d_{\ell}$) & ($\bar{L}L$)($\bar{R}R$) & $23ii$ and $ii23$ \\  
   $Q^{(8)}_{qd}$  & ($\bar{q}_i \gamma_{\mu}T^{A}q_j$) ($\bar{d}_k \gamma^{\mu}T^{A}d_{\ell}$)& & $23ii$ and $ii23$\\ 
$Q^{(1)}_{q u}$  & ($\bar{q}_i \gamma_{\mu}q_j$)($\bar{u}_k \gamma^{\mu}u_{\ell}$) &  & $23ii$ \\  
   $Q^{(8)}_{qu}$  & ($\bar{q}_i \gamma_{\mu}T^{A}q_j$) ($\bar{u}_k \gamma^{\mu}T^{A}u_{\ell}$)& &$23ii$\\ 
   \hline
   
   $Q_{d d}$  & ($\bar{d}_i \gamma_{\mu}d_j$)($\bar{d}_k \gamma^{\mu}d_{\ell}$) & ($\bar{R}R$)($\bar{R}R$) & $ii23$\\

  $Q^{(1)}_{ud}$  & ($\bar{u}_i \gamma_{\mu}u_j$)($\bar{d}_k \gamma^{\mu}d_{\ell}$) & & $ii23$\\ 
   $Q^{(8)}_{ud}$  & ($\bar{u}_i \gamma_{\mu}\tau^{I}u_j$)($\bar{d}_k \gamma^{\mu}\tau^{I}d_{\ell}$) & & $ii23$\\ 
   \hline
\end{tabular}
\caption{\small The list of four-quark SMEFT operators that can
  potentially generate an LFU ${\mathcal{O}}_{9\ell}$ at scale
  $m_b$. Here $ii=11,22$ or $33$.}
\end{center}
\label{4quarkSMEFT}
\end{table}

\subsubsection{\boldmath $C_9^{\rm U} = -1.18 \pm 0.19$}

Having identified the 5 semileptonic and 33 four-quark SMEFT operators
that have the potential to generate an LFU ${\mathcal{O}}_{9\ell}$
operator with $C_9^{\rm U} = -1.18 \pm 0.19$, we must now determine
which ones actually do this. To this end, we use the Wilson package
\cite{Aebischer:2018bkb} to perform the renormalization-group running
from the NP scale $\Lambda$ to the weak scale within SMEFT, and then
from the weak scale to $m_b$ within WET.

We begin with the semileptonic SMEFT operators and calculate the
values of $C_{9,\ell}$, $C_{10,\ell}$, $C_{9,\ell}^{\prime}$ and
$C_{10,\ell}^{\prime }$ for $\ell = e, \mu$ generated when we perform
the running from scale $\Lambda$ to $m_b$. Note that, because $Q_{\ell
  d}$ and $Q_{e d}$ involve only RH down-type quarks (see Table
\ref{tab:semilepSMEFT}), they can never generate $C_{9,\ell}$ (or
$C_{10,\ell}$). However, the other three semileptonic operators can
generate $C_{9,\ell}$. Indeed, they all generate WCs that are
universal, i.e., equal for $\ell = e, \mu$. In Table
\ref{tab:semileptoWET}, we list these operators, along with the
central value + error of the SMEFT WC that generates the desired
$C_9^{\rm U}$ and the central values of the generated WET WCs.

We now turn to the four-quark operators. The 15 operators $Q^{(1)}_{q
  d}$ and $Q^{(8)}_{qd}$ with flavour assignment $ii23$, $Q_{d d}$,
$Q^{(1)}_{ud}$, and $Q^{(1)}_{q d}$ all involve only RH down-type
quarks (see Table 4), and so cannot generate $C_{9,\ell}$ (or
$C_{10,\ell}$). However, the remaining 18 four-quark operators can
generate $C_9^{\rm U}$ and $C_{10}^{\rm U}$, but here the universal
index U means equal for $\ell = e, \mu, \tau$. In Table
\ref{tab:fourquarktoWET}, we list these operators, along with the
central value + error of the SMEFT WC that generates the desired
$C_9^{\rm U}$ and the central values of the generated WET
WCs. Although all operators can generate the right $C_9^{\rm U}$, they
are not all viable solutions of the $\bsll$ anomalies. The point is
that the preferred solution has NP contributions to $C_9^{\rm U}$, but
$C_{10}^{\rm U}$ is small. However, in Table \ref{tab:fourquarktoWET}
we see that four operators -- $[C_{qq}^{(1)}]_{2333}$,
$[C_{qq}^{(3)}]_{2333}$, $[C_{qu}^{(1)}]_{2333}$,
$[C_{qu}^{(8)}]_{2333}$ -- also generate large $C_{10}^{\rm U}$. This occurs because these operators mix strongly with $C^{(1)}_{\varphi q}$, which has a large contribution to $C_{10}$ [see Eq.~(\ref{matching})].
These operators are therefore excluded.

The upshot is that there are 3 semileptonic and 14 four-quark SMEFT
operators that generate the required $C_9^{\rm U}$ WC when run down to
the $m_b$ scale. These are therefore potential solutions to the
$\bsll$ anomalies. However, this running may also generate other
operators which may be constrained by different observables. For
example, we see that the running generates a small imaginary piece of
$C_9^{\rm U}$. While this is unimportant for $\bsll$ processes,
imaginary pieces of the WCs of other operators may be generated. In
principle, this may result in new, CP-violating contributions to other
processes, and there may be constraints on such contributions. These
other constraints are examined in the next subsubsection.

\begin{table}[H]
\begin{center}
\begin{tabular}{|cc|c|c|c|c|}
\hline 
\hline
\multicolumn{2}{|c|}{ $C_{\textrm{SMEFT}}$ ($\textrm{TeV}^{-2}$) } & $C_9^{\rm U}$ & $C_{10}^{\rm U}$ &
$C^{\prime{\rm U}}_9$ & $C^{\prime{\rm U}}_{10}$\\
\hline \hline
\rowcolor{gray!20}
$[C_{lq}^{(1)}]_{3323}$ & $-0.23\pm 0.04$ & $-1.20 -i 0.022$ & $-0.004$ & 0 & 0 \\
\rowcolor{gray!20}
$[C_{lq}^{(3)}]_{3323}$ & $-0.23\pm 0.04$ & $-1.17 - i 0.022$ & $-0.021$ & 0 & 0 \\
\hline \hline
\rowcolor{gray!20}
$[C_{qe}]_{2333}$& $-0.22\pm 0.03$ & $-1.16 - i 0.022$ & $-0.005$ & 0 & 0 \\
\hline 
\hline
\end{tabular}
\caption{Running of the semileptonic SMEFT operators: (i) central
  value + error of the WC that generates $C_9^{\rm U} = -1.18 \pm
  0.19$ and (ii) central values of predictions for $C_9^{\rm U}$,
  $C_{10}^{\rm U}$, $C^{\prime{\rm U}}_9$, $C^{\prime{\rm
      U}}_{10}$. Operators producing the desired WET WCs are
  highlighted in gray. }
\label{tab:semileptoWET}
\end{center}
\end{table}

\begin{table}[H]
\centering
\begin{tabular}{|cc|c|c|c|c|}
\hline \hline
 \multicolumn{2}{|c|}{ $C_{\textrm{SMEFT}}$ ($\textrm{TeV}^{-2}$) }  & $C_9^{\rm U}$ & $C_{10}^{\rm U}$ &
$C^{\prime{\rm U}}_9$ & $C^{\prime{\rm U}}_{10}$ \\ \hline \hline
\rowcolor{gray!20}
$[C_{qq}^{(1)}]_{1123}$ & $0.21\pm 0.03$ & $-1.17 - i 0.022$& $-0.004 $&0&0  \\
\rowcolor{gray!20}
 $[C_{qq}^{(1)}]_{2223}$ & $0.25\pm 0.04$ & $-1.18 - i 0.022$&$0.028$&0&0   \\

  $[C_{qq}^{(1)}]_{3323}$ & $2.5\pm 0.4$ & $-1.17-i 0.022$ & $104.8 + i 1.9 $& 0&0 \\
  \rowcolor{gray!20}
$[C_{qq}^{(3)}]_{1123}$ & $-0.07\pm 0.01$ & $-1.15 - i 0.022$ & $-0.019$& 0&0  \\
\rowcolor{gray!20}
 $[C_{qq}^{(3)}]_{2223}$ & $-0.091\pm 0.014$ & $-1.17 - i 0.022$ & $-0.023$& 0&0   \\

$[C_{qq}^{(3)}]_{2333}$ & $-0.13\pm 0.02$ & $-1.17 - i 0.022$ & $2.86 + i 0.05 $& 0&0   \\
\hline \hline
  \rowcolor{gray!20}
  $[C_{qd}^{(1)}]_{2311}$ & $-0.22\pm 0.03$ & $-1.18 - i 0.022 $ & $-0.005$& 0&0   \\
  \rowcolor{gray!20}
  $[C_{qd}^{(1)}]_{2322}$& $-0.22\pm 0.03$ & $-1.18 -i 0.022 $ & $-0.005 $& 0&0  \\
  \rowcolor{gray!20}
$[C_{qd}^{(1)}]_{2333}$& $-0.22\pm 0.03$ & $-1.18 -i 0.022 $ & $-0.008 $& 0&0  \\
\rowcolor{gray!20}
$[C_{qd}^{(8)}]_{2311}$ & $-3.10\pm 0.50$ & $-1.17 - i 0.022 $ & $-0.040 - i 0.0007 $& 0&0    \\
\rowcolor{gray!20}
$[C_{qd}^{(8)}]_{2322}$ & $-3.10\pm 0.50$ & $-1.17 - i 0.022$ & $-0.040 - i 0.0007$& 0&0    \\
\rowcolor{gray!20}
$[C_{qd}^{(8)}]_{2333}$ &$-3.10\pm 0.50$ & $-1.17 - i 0.022$ & $-0.041 - i 0.0007$& 0&0  \\
\rowcolor{gray!20}
$[C_{qu}^{(1)}]_{2311}$ & $0.11\pm 0.01$ & $-1.19 - i 0.022$ & $-0.004$&0& 0\\
\rowcolor{gray!20}
$[C_{qu}^{(1)}]_{2322}$ & $0.11\pm 0.01$ & $-1.19 - i 0.022$ & $-0.005$&0&0 \\

$[C_{qu}^{(1)}]_{2333}$ & $0.50\pm 0.08$ & $-1.19 - i 0.022   $ & $-20.97 - i 0.39$& 0&0   \\
\rowcolor{gray!20}
$[C_{qu}^{(8)}]_{2311}$ & $1.12\pm 0.18$ & $-1.18  -i 0.022   $ & $0.013$& 0&0   \\
\rowcolor{gray!20}
$[C_{qu}^{(8)}]_{2322}$  &$1.12\pm 0.18$ & $-1.18  -i 0.022   $ & $0.013$& 0&0 \\

$[C_{qu}^{(8)}]_{2333}$  & $18.5\pm 3.0$ & $-1.17   - i 0.022$ & $-14.83 - i 0.27$& 0&0  \\
\hline \hline
\end{tabular}
\caption{Running of the four-quark SMEFT operators: (i) central value
  + error of the WC that generates $C_9^{\rm U} = -1.18 \pm 0.19$ and
  (ii) central values of predictions for $C_9^{\rm U}$, $C_{10}^{\rm
    U}$, $C^{\prime{\rm U}}_9$, $C^{\prime{\rm U}}_{10}$. Operators
  producing the desired WET WCs are highlighted in gray. }
\label{tab:fourquarktoWET}
\end{table}

\subsection{Other constraints}

In general, through running, a given SMEFT operator will contribute to
a variety of WET operators, and there may be important constraints on
some of these. Thus, before declaring that the addition of a
particular SMEFT operator can explain the $\bsll$ anomalies, one has
to be sure that all other constraints have been taken into account. In
what follows we discuss the important constraints for each SMEFT
operator under consideration.

Note that the $\bsll$ anomalies are not the only measurements of
semileptonic $B$ decays that disagree with the SM. For many years,
there have been anomalies in the measurements of $R_{D^{(*)}}$ that
  suggest NP in $\bctaunu$ \cite{Iguro:2024hyk}. And recently, Belle
  II measured ${\mathcal B}(B^+ \to K^+ \nu {\bar\nu}$, finding a
  value almost $3\sigma$ above the SM prediction
  \cite{Belle-II:2023esi}. We do not require our SMEFT operator to
  explain these anomalies. Instead, the constraint imposed is that the
  predictions for these observables should not be larger than the
  experimental value within $2\sigma$, though it can be smaller. For
  the other constraints, we require that the prediction agree with the
  experimental value within $2\sigma$.

\subsubsection{\boldmath {$\bsnunu$}} 

We begin with $\bsnunu$ transitions. The first step is to identify the
WET operators that (i) contribute to $\bsnunu$ and (ii) can be
generated when we run the SMEFT operators down to the $m_b$ scale. The
SMEFT operators that have been identified above as having the
potential to generate the required $C_9^{\rm U}$ WC are all vector
operators. As a consequence, they will generate only vector $\bsnunu$
WET operators. These are
\beq
 \mathcal{N}   C_L (\bar s \gamma_\mu P_L b ) (\bar \nu_i \gamma^\mu (1- \gamma_5) \nu_i) + 
    \mathcal{N} C_R (\bar s \gamma_\mu P_R b ) (\bar \nu_i \gamma^\mu (1- \gamma_5) \nu_i) + h.c.,
\eeq
where $\mathcal{N}$ is defined below Eq.~(\ref{matching}).

For the first operator, the tree-level matching condition to the SMEFT operators is
\beq
    C_L  = {1\over 2 \mathcal{N}} ([C_{lq}^{(1)}]_{ii23} - [C_{lq}^{(3)}]_{ii23}) ~.
\eeq
A nonzero $[C_{lq}^{(1)}]_{3323}$ or $[C_{lq}^{(3)}]_{3323}$ are both
potential solutions to the $\bsll$ anomalies, see Table
\ref{tab:semileptoWET}. So there may be important constraints on these
WCs from $\bsnunu$.  (Note that these constraints can be evaded in a
specific model such as that of the $U_1$ leptoquark, in which
$[C_{lq}^{(1)}]_{3323} = [C_{lq}^{(3)}]_{3323}$.)  For the second
operator, the WC $C_R$ matches at tree level to $C_{ld}$. However,
$Q_{\ell d}$ does not generate the required $C_9^{\rm U}$ WC, so that
constraints on $C_{\ell d}$ are irrelevant. Of course, all the
semileptonic WCs of Table \ref{tab:semileptoWET} may generate one or
both of these WET operators through one-loop RGE running; if so, we
must compute the constraints on these SMEFT WCs.

There are three $\bsnunu$ processes that have been measured. Their
current experimental values are
\bea
   {\mathcal B}(B^+ \to K^+ \nu \bar \nu) & = & (2.3 \pm 0.5({\rm stat}) {}^{+0.5}_{-0.4}({\rm syst})) \times 10^{-5} ~, \nn \\
   {\mathcal B}(B^+ \to K^{*+}(892) \nu \bar \nu) & < & 4 \times 10^{-5} ~(90\%~{\rm C.L.}) ~, \nn \\
   {\mathcal B}(B^0\to K^0 \nu \bar \nu) & < & 2.6 \times 10^{-5} ~(90\%~{\rm C.L.}) ~.   
\label{BKnunu_expt}
\eea
We take the first measurement from Ref.~\cite{Belle-II:2023esi} and
the other two from Ref.~\cite{pdg}.

For the constraints on the SMEFT operators, we require only that the
predicted branching ratios not exceed the measured values/limits to
within $2\sigma$ (the theoretical and experimental (if applicable)
errors are combined in quadrature).

\subsubsection{\boldmath {$\bctaunu$}} 

Turning to the transition $\bctaunu$, there are again several WET
operators that contribute. However, only a single one can be generated
by the running of a dimension-6 vector SMEFT operator. It is
\beq
\mathcal{N}' \, C_V^{33} \, \left( \bar c_L \gamma^\mu b_L \right) \left(\bar\tau_L \gamma_\mu \nu_{\tau L} \right) ~, 
\eeq
where $\mathcal{N}' = {{4 G_F \over \sqrt 2} V_{cb}}$. Its tree-level
matching condition to the SMEFT operators is
\beq
    C_V^{33}  = {1\over 2 \mathcal{N}'} \, V_{2k} \, [C_{lq}^{(3)}]_{33k3}^* ~,
\eeq
where $V_{2k}$ is a CKM matrix element [see Eq.~(\ref{uquarkdef})]. In
particular, $[C_{lq}^{(3)}]_{3323}$, which generates the required
$C_9^{\rm U}$ when run down to $m_b$, also generates $\bctaunu$ at
tree level. We stress that the other semileptonic WCs of Table
\ref{tab:semileptoWET} may also generate such a transition, albeit at
loop level.

The decays ${\bar B} \to D^{(*)} \tau^- {\bar\nu}_\tau$ involve the
transition $\bctaunu$. The ratios $R_{D^{(*)}}$ are defined as
\beq
R_D = \frac{\mathcal{B}({\bar B} \to D \tau^- {\bar\nu}_\tau)}{\mathcal{B}({\bar B} \to D \ell^- {\bar\nu}_\ell)}
~~,~~~~
R_{D^*} = \frac{\mathcal{B}({\bar B} \to D^* \tau^- {\bar\nu}_\tau)}{\mathcal{B}({\bar B} \to D^* \ell^- {\bar\nu}_\ell)} ~~,~ \ell = e. \mu ~.
\eeq
These have been measured by several different experiments. The latest
world averages are \cite{Iguro:2024hyk}
\beq
R_D = 0.344 \pm 0.026 ~~,~~ R_{D^*} = 0.285 \pm 0.012 ~,
\label{RD(*)_latest}
\eeq
with a correlation of $-0.39$. Both values are larger than the SM
predictions, which are $R_D = 0.298(4)$ and $R_{D^*} = 0.254(5)$ \cite{HFLAV:2022esi}.
When the data of all the $\bctaunu$ observables are
combined, there is a $4.3\sigma$ disagreement with the SM.

As was the case with $\bsnunu$, we do not require that our SMEFT
operator explain the $\bctaunu$ anomaly. (Indeed, only
$[C_{lq}^{(3)}]_{3323}$, which generates $\bctaunu$ at tree level,
could possibly do so.) Instead, we only require that the predicted values of
$R_{D^{(*)}}$ not exceed the measured values to within $2\sigma$. That
is, these predicted values can be smaller (like the SM).

\subsubsection{\boldmath {$\Delta F=2$ observables}} 

The four-quark operators of Table \ref{tab:fourquarktoWET} violate
quark flavour by one unit at tree level. All of them have $\Delta B =
\Delta S = 1$, and those that involve up-type quarks can have
$\Delta C = 1$ due to Eq.~(\ref{uquarkdef}). This means that, at one
loop, i.e., through RGE running, they can contribute to $\Delta F=2$
processes \cite{Aebischer:2020dsw}. That is, there may be constraints
on the SMEFT operators from low-energy measurements of the real and
imaginary parts of $K^0$-${\bar K}^0$, $D^0$-${\bar D}^0$,
$B^0_d$-${\bar B}^0_d$ and $B^0_s$-${\bar B}^0_s$ mixing. This is
examined in this subsubsection.

There are a variety of $\Delta F=2$ observables that can, in principle, be used to
constrain the SMEFT operators, but we find that only two of these --
$\Delta M_s$ and $\varepsilon_K$ -- provide significant constraints. Here
we focus on these two observables.

There are scalar and vector WET operators that contribute to $\Delta
F=2$ meson-mixing processes. As usual, since we have only vector SMEFT
operators at the high-energy scale, only vector WET operators will be
generated when we run down to the scale $m_b$. These WET operators are
\begin{align}
&[O^{V,LL}_{dd}]_{ijij}  = (\bar d_i \gamma_\mu P_L d_j) (\bar d_i \gamma^\mu P_L d_j) ~, \nn\\
&[O^{V,RR}_{dd}]_{ijij}  = (\bar d_i \gamma_\mu P_R d_j) (\bar d_i \gamma^\mu P_R d_j) ~, \nn\\
&[O^{V1,LR}_{dd}]_{ijij}  = (\bar d_i \gamma_\mu P_L d_j) (\bar d_i \gamma^\mu P_R d_j ) ~, \nn\\
&[O^{V8,LR}_{dd}]_{ijij}  = (\bar d_i \gamma_\mu P_L T^A d_j) (\bar d_i \gamma^\mu P_R T^A d_j) ~,
\end{align}
where $i \ne j$. Their tree-level matching conditions to SMEFT
operators are
\begin{align}
&[C^{V,LL}_{dd}]_{ijij}  = -( [C^{(1)}_{qq}]_{ijij} + [C^{(3)}_{qq}]_{ijij}) \,, \nn\\
&[C^{V1,LR}_{dd}]_{ijij}  = - [C^{(1)}_{qd}]_{ijij} \,, \nn\\
&[C^{V8,LR}_{dd}]_{ijij}  = - [C^{(8)}_{qd}]_{ijij}\nn \,,  \\
&[C^{V,RR}_{dd}]_{ijij}   = - [C_{dd}]_{ijij}.
\end{align}
None of these SMEFT operators appear in Table
\ref{tab:fourquarktoWET}. This is good, as they would produce meson
mixing at tree level, which is excluded. On the other hand, some
operators in Table \ref{tab:fourquarktoWET} may mix at one loop with
the above operators, and will thereby contribute to $\Delta F=2$ meson
mixing.

Consider first $\Delta M_s$, which measures the real part of the
$B^0_s$-${\bar B}^0_s$ mixing amplitude. The operator
$[C_{qq}^{(1)}]_{3323}$ (see Table \ref{tab:fourquarktoWET})
contributes at tree level to $(\bar b \gamma_\mu P_L b) (\bar s
\gamma^\mu P_L b)$. But the $(\bar b \gamma_\mu P_L b)$ current can be
turned into a $(\bar s \gamma_\mu P_L b)$ current at one loop via a
box diagram similar to that describing $B^0_s$-${\bar B}^0_s$
mixing. In the same way, other $\Delta F=1$ operators in Table
\ref{tab:fourquarktoWET} may also contribute to meson mixing at one
loop, though the size of the effect depends on the exact flavour
configuration \cite{Aebischer:2020dsw}.

In order to include the $\Delta M_s$ constraint, we run the $\Delta
F=1$ SMEFT WCs down to $m_b$ and generate the $\Delta F=2$
operators. These are then converted into a prediction for $\Delta
M_s$, which is compared with its experimental value \cite{pdg}:
\bea
    (\Delta M_s)_{\textrm{exp.}} = 1.1683 (13)\times 10^{-11} ~\textrm{GeV}.
\eea
We require that the prediction and the experimental value agree to
within $2\sigma$, which places a constraint on the SMEFT WC.

The second important $\Delta F=2$ observable is $\varepsilon_K$, which
measures the imaginary part of the $K^0$-${\bar K}^0$ mixing amplitude
(i.e., indirect CP violation). As was the case for $\Delta M_s$, when
the $\Delta F=1$ SMEFT operators are run down to $m_b$, they may
generate at one loop $\Delta F=2$ operators that lead to $K^0$-${\bar
  K}^0$ mixing. Here the key point is that, even if the SMEFT WC is
real, the running can produce an imaginary part of the WET WC, as was
done with $C_9^{\rm U}$ (see Table \ref{tab:fourquarktoWET}). This may
lead to an important contribution to $\varepsilon_K$.

To place constraints on the SMEFT WCs from $\varepsilon_K$, we use
\cite{Aebischer:2020mkv}
\bea
    (\varepsilon_K)^{\textrm NP} = \kappa_{\varepsilon} \times 10^{-3} ~, \quad -0.2 \le \kappa_{\varepsilon} \le 0.2.
\eea

\subsubsection{\boldmath {CP violation in $K_L \to \pi \pi$}}

Another observable that can be used as a constraint is
$\varepsilon'/\varepsilon$, which is a measure of direct CP violation
in $K_L \to \pi \pi$ decays. This is a $\Delta F=1$ process, which
receives contributions from a variety of WET operators describing $s
\to d q {\bar q}$ ($q=u,d$) transitions.

Including isospin-breaking corrections, along with recent computations
of the matrix elements from the RBC-UKQCD lattice group
\cite{RBC:2020kdj}, the SM prediction is given by \cite{
  Buras:1999st,Buras:2020pjp}
\beq
  (\varepsilon'/\varepsilon)_{\textrm{SM}} = (13.9\pm 5.4) \times 10^{-4} ~.  
\eeq
The experimental world average of measurements from the NA48
\cite{NA48:2002tmj} and KTeV \cite{KTeV:2002qqy,Worcester:2009qt}
experiments is
\bea
     (\varepsilon'/\varepsilon)_{\textrm{exp}} = (16.3\pm 2.3) \times 10^{-4} ~,
\eea
which is consistent with the SM prediction within the errors.  

When our SMEFT operators are run down to $m_b$. they may generate some
WET $s \to d q {\bar q}$ operators. And since the running may produce
an imaginary piece for these WET WCs, there may be a contribution to
$\varepsilon'/\varepsilon$. To compute this contribution, we use the
master formula from Ref. \cite{Aebischer:2021hws}.

As a constraint, we require that the value of
$\varepsilon'/\varepsilon$ found when one includes the SM and NP
contribution agree with the experimental value within
$2\sigma$. However, we note that the error on the SM prediction is
very large (40\%). Furthermore, this error is almost entirely
theoretical, so it is only an estimate. For this reason, we will not
use the $\varepsilon'/\varepsilon$ constraint to exclude any SMEFT
operators; we will only note if there is a tension.

\subsubsection{\boldmath {High-$p_T$ searches}} 

We began our analysis by looking for SMEFT operators that contribute
to $\bsff$, where $f$ is a light SM fermion. In the literature, it has
been suggested that one can search for, or put limits on, NP in
$\bsff$ by looking at the high-$p_T$ distribution for $p p \to f {\bar
  f} X$ at the LHC.

The idea is as follows. The SM contribution to $p p \to f {\bar f} X$
comes mainly from $q {\bar q} \to f {\bar f}$ via the $s$-channel
exchange of a $\gamma$, $Z^0$ or gluon, while the NP contribution is
simply a four-fermion interaction. We can write
\beq
A = A_{SM} + A_{NP} = \frac{C_{SM}}{E^2} + \frac{C_{NP}}{\Lambda^2} ~,
\label{NPdim6}
\eeq
where, for simplicity, we take $C_{SM}$ and $C_{NP}$ to be real. This
leads to
\beq
|A|^2 = \frac{1}{E^4} \left[ C^2_{SM} + 2 C_{SM} C_{NP} \,
\frac{E^2}{\Lambda^2} + C^2_{NP} \, \frac{E^4}{\Lambda^4} \right] ~.
\eeq
At low energies ($E \ll \Lambda$), this is dominated by the SM
contribution. But as $E \to \Lambda$, the NP contributions become
increasingly important. Thus, by looking at the distribution for $p p
\to f {\bar f} X$ as a function of the $f {\bar f}$ invariant
mass$^2$, and by subtracting the SM contribution, one will be
sensitive to the NP contribution at high invariant
mass$^2$. Perhaps a signal of NP will be seen,
and if not, a limit can be placed on ${C_{NP}}/{\Lambda^2}$. This type
of analysis has been done in Refs.~\cite{Faroughy:2016osc,
  Becirevic:2024pni} ($f = \tau$), \cite{Alte:2017pme,
  Keilmann:2019cbp} ($f = q$) and \cite{Greljo:2017vvb} ($f = e,
\mu$).

The problem here is that an EFT is really only applicable at $E \ll
\Lambda$. As $E \to \Lambda$, the expansion in $1/\Lambda$ begins to
break down. For example, in Eq.~(\ref{NPdim6}) above, we included only
the dimension-6 NP contribution. But say we add a dimension-8
term. This could be a four-fermion interaction with two
derivatives. The coefficient is proportional to $1/\Lambda^4$, but by
dimensional analysis there must be a factor of $E^2$ in the numerator
(e.g., coming from the two derivatives). We then have
\bea
A &=& \frac{C_{SM}}{E^2} + \frac{C_{NP}}{\Lambda^2} + \frac{C_{NP8}}{\Lambda^4}  \nn\\
|A|^2 &=& \frac{1}{E^4} \left[ C^2_{SM} + 2 C_{SM} C_{NP} \,
\frac{E^2}{\Lambda^2} + C^2_{NP} \, \frac{E^4}{\Lambda^4} + 2 C_{SM} C_{NP8} \,
\frac{E^4}{\Lambda^4} \right] ~.
\eea
Now, for $E \simeq \Lambda$, the last contribution is as important as
the others. Even if $E < \Lambda$, the last term has the potential to
be as important as the third term, since it is possible that $C_{NP8}
> C_{NP}$. The point is that there is a certain amount of uncertainty
in this type of analysis, depending on how close $E$ is to
$\Lambda$. For this reason, we do not require our SMEFT WCs to satisfy
such constraints.

\subsubsection{SMEFT operators confront other constraints}

There are two types of SMEFT operators that can generate the correct
value of $C_9^{\rm U}$ when run down to the $m_b$ scale, semileptonic
operators (Table \ref{tab:semileptoWET}) and four-quark operators
(Table \ref{tab:fourquarktoWET}). Similarly, the other constraints
described above come in two categories, semileptonic constraints
($\bsnunu$, $R_{D^{(*)}}$) and hadronic constraints ($\Delta M_s$,
$\varepsilon_K$, $\varepsilon'/\varepsilon$). It turns out that the
constraints of a given type are important only for SMEFT operators of
the same type. That is, the semileptonic (four-quark) operators
satisfy all the hadronic (semileptonic) constraints.

In Table 7, we show the predictions of the
semileptonic SMEFT WCs of Table \ref{tab:semileptoWET} for
$\mathcal{B}(B\to K\nu\bar{\nu})$, $R_D$ and $R_{D^*}$.
$[C_{lq}^{(1)}]_{3323}$ and $[C_{lq}^{(3)}]_{3323}$ both predict
values for $\mathcal{B}(B\to K\nu\bar{\nu})$ that are much larger than
the experimental measurements [Eq.~(\ref{BKnunu_expt})], and
$[C_{lq}^{(3)}]_{3323}$ predicts values for $R_D$ and $R_{D^*}$ that
are also larger than the experimental values
[Eq.~(\ref{RD(*)_latest})]. Only $[C_{qe}]_{2333}$ satisfies all the
constraints.

\begin{table}[H]
\begin{center}
\begin{tabular}{|c|c|c|c|}
\hline 
\hline
\multicolumn{1}{|c|}{ $C_{\textrm{SMEFT}}$ } & $\mathcal{B}(B\to K\nu\bar{\nu})$ & $R_D$ &
$R_{D^*}$ \\
\hline \hline
$[C_{lq}^{(1)}]_{3323}$ & $(6.18\pm 0.88)\times 10^{-4}$ ~~($\times$) & $0.292\pm 0.007$ ~~($\surd$) & $0.242\pm 0.007$ ~~($\surd$)\\
$[C_{lq}^{(3)}]_{3323}$ & $(8.55\pm 1.04)\times 10^{-4}$ ~~($\times$) & $0.524\pm 0.013$ ~~($\times$) & 
$0.435\pm 0.010$ ~~($\times$) \\
\rowcolor{gray!20}
$[C_{qe}]_{2333}$ & $(4.38\pm 0.62)\times 10^{-6}$ ~~($\surd$) & $0.295\pm 0.006$ ~~($\surd$) & $0.245\pm 0.006$ ~~($\surd$) \\
\hline 
\hline
\end{tabular}
\caption{Semileptonic SMEFT WCs of Table \ref{tab:semileptoWET}:
  predictions for $\mathcal{B}(B\to K\nu\bar{\nu})$, $R_D$ and
  $R_{D^*}$, along with an indicator of whether the constraint is
  satisfied ($\surd$) or violated ($\times$).  Operators that satisfy all the constraints
are highlighted in gray.}
\end{center}
\label{tab:semilep_constraints}
\end{table}

In Table 8, we show the predictions of
the four-quark SMEFT WCs of Table \ref{tab:fourquarktoWET} that
generate the desired WET WCs for $\Delta M_s$, $\varepsilon_K$ and
$\varepsilon^{\prime}/\varepsilon$. Of the 14 WCs, only 6 of them --
$[C_{qq}^{(1)}]_{1123}$, $[C_{qq}^{(3)}]_{1123}$,
$[C_{qu}^{(1)}]_{2311}$, $[C_{qu}^{(1)}]_{2322}$,
$[C_{qu}^{(8)}]_{2311}$, $[C_{qu}^{(8)}]_{2322}$ -- satisfy the
constraints. Also, there is a caveat that $[C_{qq}^{(1)}]_{1123}$ has
a possible tension with $\varepsilon^{\prime}/\varepsilon$.

\begin{table}[H]
\begin{center}
\begin{tabular}{|c|c|c|c|c|}
\hline 
\hline
\multicolumn{1}{|c|}{ $C_{\textrm{SMEFT}}$ } & $\Delta M_s$ ($\times 10^{11}$ GeV) & $\kappa_{\varepsilon}$ &
$\varepsilon^{\prime}/\varepsilon$ ($\times 10^{4}$) \\
\hline \hline
\rowcolor{gray!20}
$[C_{qq}^{(1)}]_{1123}$ & $(1.15\pm 0.06) $ ~~($\surd$) &$-0.012$ ~~($\surd$)  & $38.4 $ ~~($?$)  \\
 $[C_{qq}^{(1)}]_{2223}$ & $(2.72\pm 0.10)$ ~~($\times$) & $0.11$ ~~($\surd$) & $15.8 $ ~~($\surd$) \\
\rowcolor{gray!20}
$[C_{qq}^{(3)}]_{1123}$  & $(1.16\pm 0.06)$ ~~($\surd$) & $-0.005$ ~~($\surd$) & $23.1 $ ~~($\surd$) \\
 $[C_{qq}^{(3)}]_{2223}$  & $(0.59\pm 0.05)$ ~~($\times$)  & $-0.04 $~~($\surd$)  & $ 17.8 $ ~~($\surd$) \\
\hline \hline
  $[C_{qd}^{(1)}]_{2311}$ & $(1.16\pm 0.07)$ ~~($\surd$) & $-0.75 $ ~~($\times$) & $13.9 $  ~~($\surd$) \\
  $[C_{qd}^{(1)}]_{2322}$ & $(1.55\pm 0.07)$ ~~($\times$)  & $0.75 $ ~~($\times$) & $13.9 $ ~~($\surd$) \\
$[C_{qd}^{(1)}]_{2333}$ & $(0.76\pm 0.06)$ ~~($\times$)  & $0.0$ ~~($\surd$) & $13.9 $  ~~($\surd$)  \\
$[C_{qd}^{(8)}]_{2311}$  & $(1.16\pm 0.06)$ ~~($\surd$) & $-15.0 $ ~~($\times$) & $12.8 $  ~~($\surd$)  \\
$[C_{qd}^{(8)}]_{2322}$ & $(1.18\pm 0.05)$ ~~($\surd$)  & $14.3 $ ~~($\times$) & $12.8 $  ~~($\surd$)  \\
$[C_{qd}^{(8)}]_{2333}$  & $(10.6\pm 0.5)$ ~~($\times$)  & $-0.001 $ ~~($\surd$) & $12.8 $ ~~($\surd$) \\
\rowcolor{gray!20}
$[C_{qu}^{(1)}]_{2311}$ & $(1.15\pm 0.06)$ ~~($\surd$)  & $0.0$ ~~($\surd$) & $13.9 $  ~~($\surd$) \\
\rowcolor{gray!20}
$[C_{qu}^{(1)}]_{2322}$  & $(1.16\pm 0.06)$ ~~($\surd$) & $0.0$ ~~($\surd$) & $13.9 $  ~~($\surd$) \\
\rowcolor{gray!20}
$[C_{qu}^{(8)}]_{2311}$  & $(1.15\pm 0.06)$ ~~($\surd$) & $0.0002 $ ~~($\surd$) & $13.9 $  ~~($\surd$) \\
\rowcolor{gray!20}
$[C_{qu}^{(8)}]_{2322}$   &  $(1.15\pm 0.06)$ ~~($\surd$) & $-0.0003$ ~~($\surd$)  &  $13.9 $ ~~($\surd$) \\
\hline\hline
\end{tabular}
\caption{Four-quark SMEFT WCs of Table \ref{tab:fourquarktoWET} that
  generate the desired WET WCs: predictions for $\Delta M_s$,
  $\kappa_{\varepsilon}$ and $\varepsilon'/\varepsilon$,along with an
  indicator of whether the constraint is satisfied ($\surd$) or
  violated ($\times$), or if there is a tension ($?$). Operators that
  satisfy all the constraints are highlighted in gray.}
\end{center}
\label{tab:fourquark_constraints}
\end{table}

\section{\boldmath SMEFT and $\btopik$}
\label{SimExp}

In Sec.~\ref{SecpiK}, we saw that the $\btopik$ amplitudes can be
expressed in terms of diagrams. When the ratio $|C'/T'|$ is fixed to
0.2 (its preferred theoretical value) and a fit to the data is
performed, we find that there is a $2.8\sigma$ discrepancy with the
SM. This is the $\btopik$ puzzle.

We also saw that all NP contributions to $\btopik$ can be combined
into three distinct matrix elements, each with its own weak phase. If
all these phases are different, then there are too many unknown
parameters to do a fit. However, if all three NP matrix elements are
generated when a single SMEFT operator is run down to the scale $m_b$,
then there will be only two NP parameters, the magnitude and the phase
of the SMEFT WC. In this case, a fit can be performed, and we can see
if the addition of the NP produces a good fit. 

In the previous section, we found seven SMEFT operators that, when
run down to the scale $m_b$, solve the $\bsll$ anomalies by generating
$C_9^{\rm U} = -1.18 \pm 0.19$, and are also consistent with all the
other constraints. In this section, we examine whether any of these
can also generate the right values of the NP $\btopik$ matrix elements
to also solve the $\btopik$ puzzle.

\subsection{New physics}

Reminder: the three NP matrix elements that appear in the $\btopik$
amplitudes are $\ApNPcomb \equiv - \ApNPuph + \ApNPdph$, $\ApNPCuph$,
and $\ApNPCdph$, where the individual contributions are defined in
Eq.~(\ref{eq: amplitudes}) and are repeated here for convenience:
\bea
\label{eq: amplitudes2}
\sum \bra{\pi K} {\bar s}_\alpha \Gamma_i b_\alpha \, {\bar q}_\beta \Gamma_j q_\beta \ket{B} &\equiv& \ApNPqph
~, \nn\\
\sum \bra{\pi K} {\bar s}_\alpha \Gamma_i b_\beta \, {\bar q}_\beta \Gamma_j q_\alpha \ket{B} &\equiv& \ApNPCqph
~~,~~ q = u,d ~.
\eea
In the above, the $\bsqq$ WET operators can have any Lorentz
structure. But in our case, since they are generated from the running
of vector SMEFT operators, only the vector WET operators are
relevant. Four of these are color-allowed:
\bea
\label{eq:WETBKPi1}
C^q_{V_{LL}} \, \Npp (\bar{s}_L\gamma_{\mu}b_L) (\bar{q}_L\gamma^{\mu}q_L) & ~,~~ &
C^q_{V_{LR}} \, \Npp (\bar{s}_L\gamma_{\mu}b_L) (\bar{q}_R\gamma^{\mu}q_R) ~, \nn\\
C^q_{V_{RL}} \, \Npp (\bar{s}_R\gamma_{\mu}b_R) (\bar{q}_L\gamma^{\mu}q_L) & ~,~~ &
C^q_{V_{RR}} \, \Npp (\bar{s}_R\gamma_{\mu}b_R) (\bar{q}_R\gamma^{\mu}q_R) ~,
\eea
where $\Npp = {{4 G_F \over \sqrt 2} V_{tb} V^*_{ts}}$. Four are
color-suppressed:
\bea
\label{eq:WETBKPi2}
C^{q,C}_{V_{LL}} \, \Npp (\bar{s}^{\alpha}_L\gamma_{\mu}b^{\beta}_L)
(\bar{q}^{\beta}_L\gamma^{\mu}q^{\alpha}_L) & ~,~~ &
C^{q,C}_{V_{LR}} \, \Npp (\bar{s}^{\alpha}_L\gamma_{\mu}b^{\beta}_L)
(\bar{q}^{\beta}_R\gamma^{\mu}q^{\alpha}_R) ~, \nn\\
C^{q,C}_{V_{RL}} \, \Npp (\bar{s}^{\alpha}_R\gamma_{\mu}b^{\beta}_R)
(\bar{q}^{\beta}_L\gamma^{\mu}q^{\alpha}_L) & ~,~~ &
C^{q,C}_{V_{RR}} \, \Npp (\bar{s}^{\alpha}_R\gamma_{\mu}b^{\beta}_R)
(\bar{q}^{\beta}_R\gamma^{\mu}q^{\alpha}_R) ~.
\eea

The matrix elements in Eq.~(\ref{eq: amplitudes2}) have been computed
for vector WET operators in Ref.~\cite{Beneke:2001ev} using QCD
factorization. With these results, we can write the three NP $\btopik$
matrix elements in terms of the WET WCs of Eqs.~(\ref{eq:WETBKPi1},
\ref{eq:WETBKPi2}):
\bea
    A^{\prime C,d} e^{i\phi^{\prime C}_u} &=&  \lambda_t \, C^{d,C}_V A_{\pi K} ~, \nonumber \\
    A^{\prime C,u} e^{i\phi^{\prime C}_d} &=&  \lambda_t \, C^{u,C}_V A_{\pi K} ~, \nonumber \\
    A^{\prime comb} e^{i\phi^\prime} &=&  \lambda_t \, C^{ud}_V A_{K \pi }\, ~, 
\label{NP_ACconversion}
\eea
where $\lambda_t = V^*_{tb} V_{ts}  $. Here, 
\bea
C^{d,C}_V    &=&  C^{d,C}_{V_{LL}} + r_{\chi}^K C^{d,C}_{V_{LR}} - C^{d,C}_{V_{RR}} - r_{\chi}^K C^{d,C}_{V_{RL}} ~, \nn\\
C^{u,C}_V  &=&  C^{u,C}_{V_{LL}} + r^K_{\chi}C^{u,C}_{V_{LR}} -C^{u,C}_{V_{RR}} - r_{\chi}^K C^{u,C}_{V_{RL}}  ~, \nn\\
 C^{ud}_V  &=&   \left(-C^u_{V_{LL}} +C^d_{V_{LL}}+C^u_{V_{LR}}-C^d_{V_{LR}}\right) \nn\\
        && \hskip5truemm +~\left(C^u_{V_{RR}} -C^d_{V_{RR}}-C^u_{V_{RL}}+C^d_{V_{RL}}\right) ~, 
\label{CNPdefs}
\eea 
with
\beq
    r^{K}_{\chi}(\mu) = \frac{2m^2_K}{m_b(\mu)(m_q(\mu)+m_s(\mu))} ~,
\eeq
which is formally of $O(1/m_b)$, but is numerically close to unity. Also,
\bea
    A_{\pi K} &=& i\frac{G_F}{\sqrt{2}}(m^2_B-m^2_{\pi})F_0^{B\to \pi }(m^2_K) f_K ~, \nonumber\\
    A_{K\pi} &=& i\frac{G_F}{\sqrt{2}}(m^2_B-m^2_{K})F_0^{B\to K }(m^2_{\pi}) f_{\pi} ~,
\eea
where $F^{B\to M}_0(q^2)$ are semileptonic form factors, and $f_\pi$
($f_K$) is the pion (kaon) decay constant.

In order to illustrate the effect of NP, we (i) add the three NP
matrix elements to the $\btopik$ amplitudes [see
  Eqs.~(\ref{BpiKNPamps}) and (\ref{NP_ACconversion})], (ii) assume
that all three have the same NP weak phase, and (iii) redo the fit to
the $\btopik$ data. The results are shown in Table
\ref{tab:NPfitCT0.2}. We see that, with the addition of NP, one
obtains a good fit to the data. These results suggest that the
$\ApNPCu$ NP amplitude is the most important, with $\ApNPCd$ being
smaller, and ${\cal A}^{\prime, comb}$ irrelevant (it is consistent
with zero).

\begin{table}[h!]
    \centering
    \begin{tabular}{|c|c|}
\hline
\multicolumn{2}{|c|}{$\chi^2_{\rm min}/{\rm d.o.f.} = 0.30/1$,}\\
\multicolumn{2}{|c|}{p-value $= 0.58$}\\
    \hline
     parameter    & best fit value  \\
     \hline
$\gamma$ & $(66.46 \pm 3.44)^\circ$ \\
    \hline
$\beta$ & $(22.12 \pm 0.69)^\circ$ \\
    \hline
    $|T^{\prime}|$ & $-5.2\pm 1.2$\\
    \hline
    $|P^{\prime}_{tc}|$ & $51.8\pm 0.5$\\
    \hline
    $C^{d,C}_V$ & $0.012\pm 0.007$\\
    \hline
    $C^{u,C}_V$ & $0.074\pm 0.017$\\
    \hline
    $C^{ud}_V$ & $0.0\pm 1.4$\\
    \hline
    $\delta_{P^{\prime}_{tc}}$ & $(198.2 \pm 4.0)^\circ$ \\
    \hline
    $\delta_{C^{\prime}}$ & $(-28.7 \pm 34.4)^\circ$\\
    \hline
    $\phi$ & $(1.95 \pm 0.57)^\circ$\\
    \hline
    \end{tabular}
\caption{$\chi^2_{\rm min}/{\rm d.o.f.}$ and best-fit values of
  unknown parameters in amplitudes of Eq.~(\ref{BpiKNPamps}) [see also
    Eq.~(\ref{NP_ACconversion})], with the same NP weak phase $\phi$
  for all NP diagrams.  Constraints: $\btopik$ data, measurements of
  $\beta$ and $\gamma$, theoretical inputs $|C'/T'| = 0.2$, $P'_{uc} =
  0$. }
\label{tab:NPfitCT0.2}
\end{table}

\subsection{Simultaneous explanations}

We have identified seven SMEFT operators that can solve the $\bsll$
anomalies by generating $C_9^{\rm U} = -1.18 \pm 0.19$ when run down
to the scale $m_b$. This running will also generate the WET operators
of Eqs.~(\ref{eq:WETBKPi1}, \ref{eq:WETBKPi2}). Given the values of
the WET WCs, we can compute the real and imaginary values of
$C_V^{d,C}$, $C_V^{u,C}$ and $C_V^{ud}$ [Eq.~(\ref{CNPdefs})]. We can
then perform a fit to the $\btopik$ data and see if the fit is good.

We follow this procedure for each of the seven candidate SMEFT
operators. The results can be found in Table 10. We see that, in fact,
none of the SMEFT operators produce a good fit. The fits of two of
them -- $[C_{qq}^{(1)}]_{1123}$ and $[C_{qu}^{(8)}]_{2322}$ -- have
p-values of 0.12, which is passable, but we are looking for stronger
explanations of the $\btopik$ data.

\begin{table}[H]
\begin{center}
\resizebox{0.9\textwidth}{!}{
\begin{tabular}{|c|c|c|c|c|c|}
\hline \hline
\multicolumn{1}{|c|}{ $C_{\textrm{SMEFT}}$ } & $C_V^{d,C}$ & $C_V^{u,C}$ & $C_V^{ud}$ & $\chi^2_{\rm min}/{\rm d.o.f.}$ &
p-value \\
\hline \hline
$[C_{qe}]_{2333}$ &  0 & 0  & 0.0001& 16.7 & 0.005 \\
\hline \hline
$[C_{qq}^{(1)}]_{1123}$ & $0.056+0.001i$ & $0.056+0.001i$ &$-0.001$ & 8.8/5& 0.12 \\
$[C_{qq}^{(3)}]_{1123}$ & $-0.020-0.0004i$ & $0.022+0.0004i$ & $0.123+0.002i$ & 15.1/5 & 0.01 \\
\hline \hline
$[C_{qu}^{(1)}]_{2311}$ &$0.0016 + 0.00003i$ &$-0.043-0.001i$ & $-0.071-0.0013i$ &14.7/5 & 0.01\\
$[C_{qu}^{(1)}]_{2322}$  &$0.0016 + 0.00003i$ & $0.0017 + 0.00003i$& $-0.0001$& 16.5/5& 0.005 \\
$[C_{qu}^{(8)}]_{2311}$  & $0.056 + 0.001i$&$-0.915-0.017i$ & $0.120+0.002i$&65.0/5 & 0\\
$[C_{qu}^{(8)}]_{2322}$ & $0.056 + 0.001$ & $0.056 + 0.001$& 0& 8.8/5 & 0.12\\
\hline\hline
\end{tabular}
}
\caption{Candidate SMEFT WCs: predictions for the NP $\btopik$
  parameters $C_V^{d,C}$, $C_V^{u,C}$ and $C_V^{ud}$, along with the
  result of the $\btopik$ fit in terms of $\chi^2_{\rm min}/{\rm
    d.o.f.}$ and the p-value. Operators that produce a good fit to the
  $\btopik$ data are highlighted in gray.}
\end{center}
\label{tab:SMEFTBpiKfit}
\end{table}

It is clear what is going on here. When the $\btopik$ fit was
performed including NP, a good fit was found, see Table
\ref{tab:NPfitCT0.2}. However, a small, nonzero NP weak phase was
required. But the SMEFT WCs in Table 10 are all real, which leads to
$\btopik$ fits that are passable at best.

To correct this problem, we add a small NP weak phase to the SMEFT
WCs.  The results are shown in Table 11. For $\bsll$, nothing has
changed -- all SMEFT WCs still generate the desired value of $C_9^{\rm
  U}$. But there is a marked difference in the $\btopik$ fits: now
three WCs -- $[C^{(1)}_{qq}]_{1123}$, $[C^{(1)}_{qu}]_{2311}$ and
$[C^{(8)}_{qu}]_{2322}$ -- produce good $\btopik$ fits. A fourth WC,
$[C^{(3)}_{qq}]_{1123}$, has a passable fit. Because the added NP weak
phase is small, the constraints from the CP-violating observables
$\kappa_{\epsilon}$, $\varepsilon'/\varepsilon$ and $\beta_s$ (the
imaginary part of $B^0_s$-${\bar B}^0_s$ mixing) are satisfied.
(However, note that $[C_{qq}^{(1)}]_{1123}$ still has a possible
tension with $\varepsilon^{\prime}/\varepsilon$, see Table 8.)

\begin{table}
\begin{center}
\begin{tabular}{|cc|c|c|}
\hline
\hline
\multicolumn{2}{|c|}{ $C_{\textrm{SMEFT}}$ ($\textrm{TeV}^{-2}$) }  & $C_9^{\rm U}$ & $C_{10}^{\rm U}$ \\
\hline 
\hline 
$[C_{qe}]_{2333}$ & $-0.22e^{0.10i}$ & $-1.15 -  0.14i$ & $-0.005-0.0006i$ \\
\hline 
\hline 
$[C^{(1)}_{qq}]_{1123}$ & $0.21e^{0.05i}$ & $-1.17-0.08i$ & $-0.004-0.0003i$  \\
 \hline
  $[C^{(3)}_{qq}]_{1123}$ &$-0.07e^{0.10i}$ &$-1.14-0.14i$& $-0.019-0.002i$ \\
  \hline
  \hline
   $[C^{(1)}_{qu}]_{2311}$ & $0.11e^{0.10i}$& $-1.18-0.14i$ & $-0.004-0.0005i$ \\
   \hline
   $[C_{qu}^{(1)}]_{2322}$ & $0.11e^{0.20i}$ & $-1.16-0.26i$ & $-0.005-0.001i$ \\
  \hline
  $[C_{qu}^{(8)}]_{2311}$ & $1.12e^{0.10i}$ & $-1.17  - 0.14i   $ & $0.013-0.001i$ \\
  \hline
  $[C^{(8)}_{qu}]_{2322}$ & $1.12e^{0.05i}$ & $-1.18-0.08i$ & $0.012+0.0008i$ \\
 \hline
\end{tabular}
\begin{tabular}{|cc|c|c|c|c|c|}
\hline
\multicolumn{2}{|c|}{ $C_{\textrm{SMEFT}}$ ($\textrm{TeV}^{-2}$) }  &$C^{d,C}_{V}$ & $C^{u,C}_{V}$ & $C^{ud}_{V}$ & $\chi^2_{\rm min}/{\rm d.o.f.}$& p-value\\
\hline
\hline 
$[C_{qe}]_{2333}$ & $-0.22e^{0.10i}$ & 0 & 0  & $0.0001e^{-3.02i}$& 16.5/5 & 0.005 \\
\hline \hline
\rowcolor{gray!20}
 $[C^{(1)}_{qq}]_{1123}$ & $0.21e^{0.05i}$ & $0.056e^{-0.0001i}$ &$0.056e^{0.07i}$ & $0.001e^{2.78i}$ &  0.43/5 & 0.99  \\
 \hline
  $[C^{(3)}_{qq}]_{1123}$ &$-0.07e^{0.10i}$ & $0.02e^{3.12i}$&$0.022e^{0.115i}$ & $0.123e^{0.116i}$& 8.8/5& 0.12\\
  \hline
  \hline
\rowcolor{gray!20}
   $[C^{(1)}_{qu}]_{2311}$ & $0.11e^{0.10i}$ & $0.0016e^{0.03i}$ & $0.043e^{-3.02i}$ & $0.071e^{-3.02i}$ & $2.67/5$ & 0.75 \\
  \hline
$[C_{qu}^{(1)}]_{2322}$ & $0.11e^{0.20i}$ &$0.0016e^{0.042i}$ & $0.0017e^{0.22i}$& $0.0001e^{-2.97}$& 14.5/5&  0.01\\
  \hline
$[C_{qu}^{(8)}]_{2311}$ & $1.12e^{0.10i}$& $0.056e^{0.057i}$&$0.915e^{-3.02}$ & $0.120e^{0.12i}$&62.4/5 & 0.0\\
\rowcolor{gray!20}
  $[C^{(8)}_{qu}]_{2322}$ & $1.12e^{0.05i}$ & $0.056e^{0.038i}$ & $0.056e^{0.069i}$ & $0.00016e^{1.24i}$ & $3.65/5$
  &  0.60\\
 \hline
 \hline
 \end{tabular}
\end{center}
\caption{Candidate SMEFT WCs: predictions for (i) $C_9^{\rm U}$ and
  $C_{10}^{\rm U}$ of $\bsll$ (upper table), and (ii) the NP $\btopik$
  parameters $C_V^{d,C}$, $C_V^{u,C}$ and $C_V^{ud}$, along with the
  result of the $\btopik$ fit in terms of $\chi^2_{\rm min}/{\rm
    d.o.f.}$ and the p-value (lower table). Operators that produce a
    good fit to the $\btopik$ data are highlighted in gray.}
 \end{table}

\subsection{Discussion}

We have found three four-quark SMEFT operators which can provide a combined explanation of the $\bsll$ anomalies and the $\btopik$ puzzle. One obvious question is: is it possible to
distinguish these scenarios? Each SMEFT operator contributes at tree
level to a set of four-quark WET transitions, and these sets are not
the same for all three operators, so the answer is potentially yes.

To be specific, from Table 4 we see that the operators (i)
$[Q^{(1)}_{qq}]_{1123}$, (ii) $[Q^{(1)}_{qu}]_{2311}$ and (iii)
$[Q^{(8)}_{qu}]_{2322}$ respectively include (i) colour-allowed
$\bsdd$, $\bsuu$, $\bscc$, $\bsuc$ and $\bscu$ transitions, (ii) only
colour-allowed $\bsuu$ transitions, and (iii) both colour-allowed and
colour-suppressed $\bscc$ transitions. All operators also contribute
to processes in which the $\btos$ is replaced by $\ttoc$. The point is
that the three operators will affect different types of hadronic $B$
decays.  The measurements of a variety of such decays may reveal which
ones are affected by NP or not, which will allow us to distinguish the
three solutions.

Another question one might ask is: what model of NP can explain
$\bsll$ and $\btopik$? This is complicated. We have found solutions
with a single SMEFT operator. But realistic models typically contain
many SMEFT operators, so our solutions only provide a starting point.
The full NP model would presumably include one of our SMEFT operators,
but then it would be necessary to compute the contributions of the
other SMEFT operators to $\bsll$ and $\btopik$, in order to see if the
model still provides an explanation of the two anomalies.

One important ingredient in our analysis was the use of both semileptonic
and four-quark operators within SMEFT. As we have stressed, due to RGE running, each type
of SMEFT operator can potentially generate the other type of WET
operator at one loop. This was key in identifying the SMEFT operators that could generate
the desired value of $C_9^{\rm U}$ in $\bsll$. In the literature, most
analyses generally focus on the semileptonic WET operator $\bstautau$
(for example, see Ref.~\cite{Alguero:2022wkd}). However, we showed
that there are more four-quark SMEFT operators (i.e., $\bsqq$ WET
operators) that can do this.

We also found that, while some hadronic SMEFT operators can explain both the
semileptonic $\bsll$ anomalies and the hadronic $\btopik$ puzzle, the semileptonic SMEFT operator that  explains $\bsll$ cannot also
explain $\btopik$. In this particular case, the
RGE running of the semileptonic operator, which involves only
electroweak gauge bosons, did generate the $\bsqq$, $q=u,d$ WET operators, but the WCs were not
sufficiently large. The four-quark SMEFT operators
did not have this problem because their RGE running involves both gluons and the Higgs in addition to electroweak gauge bosons.
This suggests that it may be difficult to explain hadronic anomalies with semileptonic SMEFT operators. However, this cannot be concluded definitively, so it is worthwhile to continue to explore this possibility.

Finally, we note that this is the first time anyone has attempted to explain the $\btopik$ puzzle within SMEFT.
In our treatment of the $\btopik$ puzzle, we minimized the theoretical
hadronic input. We fit to the SM parameters (no form-factor
calculations were needed), and we treated the NP effects with
factorization. (While one must worry about non-factorizable effects in
performing SM calculations, these are just second-order corrections
when dealing with NP.)

\section{Conclusions}

Despite its success in explaining almost all experimental data to date, we know that the standard model is not complete -- there must be physics beyond the SM. Since no new particles have been seen at the LHC, we also know that this new physics, whatever it is, must be heavy, with masses greater than $O({\rm TeV})$. When these NP particles are integrated out, one obtains the SMEFT, which includes the (dimension-4) SM and higher-order non-SM operators.

Because the NP is very heavy, it is likely that the first signs of NP will be indirect: the measured value of an observable in a low-energy process disagrees with the prediction of the SM. Whenever such an anomaly is seen, we want to know what kind of NP could account for it. In SMEFT language, this amounts to asking which SMEFT operator(s) can generate the low-energy WET operator that describes the observable.

If the anomaly is in a semileptonic or hadronic process, the first thought is to look for the semileptonic (two quarks and two leptons) or hadronic (four quarks) SMEFT operators that contain the desired WET operator. One thing we have stressed in this paper is that this is not enough. When one uses the renormalization-group equations to evolve the SMEFT Hamiltonian down to low energies, due to operator mixing semileptonic (hadronic) SMEFT operators can generate hadronic (semileptonic) WET operators at one loop. The point is that, if one wants to find an explanation of any low-energy anomaly, or combination of anomalies, within SMEFT, one must
(i) identify the candidate semileptonic and four-quark SMEFT operators, (ii) run them down to low energy with the RGEs, (iii) generate the required WET operators with the correct WCs, and (iv) check that all other constraints are satisfied.
 
In this paper, we have illustrated this method by applying it to two anomalies in the $B$ system. We have found three four-quark SMEFT operators which, when run down to the scale $m_b$, can simultaneously explain the semileptonic $\bsll$ anomalies and the hadronic $\btopik$ puzzle. A key ingredient in our analysis was considering both semileptonic and hadronic SMEFT operators. Note that, while different NP scenarios have been proposed to explain each of these anomalies, this is the first time a combined explanation has been found.

\bigskip\bigskip
\noindent
{\bf Acknowledgements:} A.D.\ thanks the SLAC National Accelerator Laboratory and the Santa Cruz Institute of Particle Physics
for their hospitality during the completion of this work. This work was financially supported by the
U.S. National Science Foundation under Grant No.\ PHY-2309937 (AD), by
the U.S. Department of Energy through the Los Alamos National
Laboratory and by the Laboratory Directed Research and Development
program of Los Alamos National Laboratory under project numbers
20220706PRD1 and 20240078DR (JK), and by NSERC of Canada (DL). Los
Alamos National Laboratory is operated by Triad National Security,
LLC, for the National Nuclear Security Administration of
U.S. Department of Energy (Contract No.  89233218CNA000001).


\newpage
\begin{thebibliography}{99}


\bibitem{Buchmuller:1985jz}
W.~Buchmuller and D.~Wyler,
``Effective Lagrangian Analysis of New Interactions and Flavor Conservation,''
Nucl.\ Phys.\ B \textbf{268}, 621-653 (1986)

\bibitem{Grzadkowski:2010es}
B.~Grzadkowski, M.~Iskrzynski, M.~Misiak and J.~Rosiek,
``Dimension-Six Terms in the Standard Model Lagrangian,''
JHEP \textbf{10}, 085 (2010)
[arXiv:1008.4884 [hep-ph]].

\bibitem{Brivio:2017vri} For a review, see
I.~Brivio and M.~Trott,
``The Standard Model as an Effective Field Theory,''
Phys. Rept. \textbf{793}, 1-98 (2019)
[arXiv:1706.08945 [hep-ph]].

\bibitem{Buras:2014fpa}
A.~J.~Buras, J.~Girrbach-Noe, C.~Niehoff and D.~M.~Straub,
``$ B\to {K}^{\left(\ast \right)}\nu \overline{\nu} $ decays in the Standard Model and beyond,''
JHEP \textbf{02}, 184 (2015)
[arXiv:1409.4557 [hep-ph]].

\bibitem{Kumar:2021yod}
J.~Kumar,
``Renormalization group improved implications of semileptonic operators in SMEFT,''
JHEP \textbf{01}, 107 (2022)
[arXiv:2107.13005 [hep-ph]].

\bibitem{Aebischer:2020dsw}
J.~Aebischer, C.~Bobeth, A.~J.~Buras and J.~Kumar,
``SMEFT ATLAS of $\Delta$F = 2 transitions,''
JHEP \textbf{12}, 187 (2020)
[arXiv:2009.07276 [hep-ph]].

\bibitem{Aebischer:2020lsx}
J.~Aebischer and J.~Kumar,
``Flavour violating effects of Yukawa running in SMEFT,''
JHEP \textbf{09}, 187 (2020)
[arXiv:2005.12283 [hep-ph]].

\bibitem{Aebischer:2020mkv}
J.~Aebischer, A.~J.~Buras and J.~Kumar,
``Another SMEFT story: $Z'$ facing new results on
$\varepsilon'/\varepsilon$, $\Delta M_{K}$ and $K \to \pi \nu \bar{\nu}$,''
JHEP \textbf{12}, 097 (2020)
[arXiv:2006.01138 [hep-ph]].

\bibitem{Bobeth:2017ecx}
C.~Bobeth and A.~J.~Buras,
``Leptoquarks meet $\varepsilon'/\varepsilon$ and rare Kaon processes,''
JHEP \textbf{02}, 101 (2018)
[arXiv:1712.01295 [hep-ph]].

\bibitem{Bobeth:2017xry}
C.~Bobeth, A.~J.~Buras, A.~Celis and M.~Jung,
``Yukawa enhancement of $Z$-mediated new physics in $\Delta S = 2$ and $\Delta B = 2$ processes,''
JHEP \textbf{07}, 124 (2017)
[arXiv:1703.04753 [hep-ph]].

\bibitem{Alok:2021ydy}
A.~K.~Alok, A.~Dighe, S.~Gangal and J.~Kumar,
``Leptonic operators for the Cabbibo angle anomaly with SMEFT RG evolution,''
Phys. Rev. D \textbf{108}, no.11, 113005 (2023)
[arXiv:2108.05614 [hep-ph]].

\bibitem{Cirigliano:2023nol}
V.~Cirigliano, W.~Dekens, J.~de Vries, E.~Mereghetti and T.~Tong,
``Anomalies in global SMEFT analyses. A case study of first-row CKM unitarity,''
JHEP \textbf{03}, 033 (2024)
[arXiv:2311.00021 [hep-ph]].

\bibitem{London:2021lfn}
For example, see D.~London and J.~Matias,
``$B$ Flavour Anomalies: 2021 Theoretical Status Report,''
Ann.\ Rev.\ Nucl.\ Part.\ Sci.\ \textbf{72}, 37-68 (2022)
[arXiv:2110.13270 [hep-ph]].

\bibitem{Beaudry:2017gtw}
N.~B.~Beaudry, A.~Datta, D.~London, A.~Rashed and J.~S.~Roux,
``The $B \to \pi K$ puzzle revisited,''
JHEP \textbf{01}, 074 (2018)
[arXiv:1709.07142 [hep-ph]].

\bibitem{Bhattacharya:2021shk}
B.~Bhattacharya, A.~Datta, D.~Marfatia, S.~Nandi and J.~Waite,
``Axion-like particles resolve the $B \to \pi K$ and $g-2$ anomalies,''
Phys.\ Rev.\ D \textbf{104}, no.5, L051701 (2021)
[arXiv:2104.03947 [hep-ph]].

\bibitem{Bhattacharya:2022akr}
B.~Bhattacharya, S.~Kumbhakar, D.~London and N.~Payot,
``U-spin puzzle in B decays,''
Phys. Rev. D \textbf{107}, no.1, L011505 (2023)
[arXiv:2211.06994 [hep-ph]].

\bibitem{Amhis:2022hpm}
Y.~Amhis, Y.~Grossman and Y.~Nir,
``The branching fraction of $ {B}_s^0\to {K}^0{\overline{K}}^0 $: three puzzles,''
JHEP \textbf{02}, 113 (2023)
[arXiv:2212.03874 [hep-ph]].

\bibitem{Biswas:2023pyw}
A.~Biswas, S.~Descotes-Genon, J.~Matias and G.~Tetlalmatzi-Xolocotzi,
``A new puzzle in non-leptonic B decays,''
JHEP \textbf{06}, 108 (2023)
[arXiv:2301.10542 [hep-ph]].

\bibitem{Bhattacharya:2014wla}
B.~Bhattacharya, A.~Datta, D.~London and S.~Shivashankara,
``Simultaneous Explanation of the $R_K$ and $R(D^{(*)})$ Puzzles,''
Phys.\ Lett.\ B \textbf{742}, 370-374 (2015)
[arXiv:1412.7164 [hep-ph]].

\bibitem{Greljo:2015mma}
A.~Greljo, G.~Isidori and D.~Marzocca,
``On the breaking of Lepton Flavor Universality in B decays,''
JHEP \textbf{07}, 142 (2015)
[arXiv:1506.01705 [hep-ph]].

\bibitem{Calibbi:2015kma}
L.~Calibbi, A.~Crivellin and T.~Ota,
``Effective Field Theory Approach to $b\to s\ell\ell^{(')}$, $B\to  K^{(*)}\nu\overline{\nu}$ and $B\to D^{(*)}\tau\nu$ with Third  Generation Couplings,''
Phys.\ Rev.\ Lett.\ \textbf{115}, 181801 (2015)
[arXiv:1506.02661 [hep-ph]].

\bibitem{Barbieri:2015yvd}
R.~Barbieri, G.~Isidori, A.~Pattori and F.~Senia,
``Anomalies in $B$-decays and $U(2)$ flavour symmetry,''
Eur.\ Phys.\ J. C \textbf{76}, no.2, 67 (2016)
[arXiv:1512.01560 [hep-ph]].

\bibitem{Boucenna:2016qad}
S.~M.~Boucenna, A.~Celis, J.~Fuentes-Martin, A.~Vicente and J.~Virto,
``Phenomenology of an $SU(2) \times SU(2) \times U(1)$ model with lepton-flavour non-universality,''
JHEP \textbf{12}, 059 (2016)
[arXiv:1608.01349 [hep-ph]].

\bibitem{Bhattacharya:2016mcc}
B.~Bhattacharya, A.~Datta, J.~P.~Gu\'evin, D.~London and R.~Watanabe,
``Simultaneous Explanation of the $R_K$ and $R_{D^{(*)}}$ Puzzles: a Model Analysis,''
JHEP \textbf{01}, 015 (2017)
[arXiv:1609.09078 [hep-ph]].

\bibitem{Crivellin:2017zlb}
A.~Crivellin, D.~M\"uller and T.~Ota,
``Simultaneous explanation of $R(D^{(*)})$ and $\bsmumu$: the last scalar leptoquarks standing,''
JHEP \textbf{09}, 040 (2017)
[arXiv:1703.09226 [hep-ph]].

\bibitem{Buttazzo:2017ixm}
D.~Buttazzo, A.~Greljo, G.~Isidori and D.~Marzocca,
``B-physics anomalies: a guide to combined explanations,''
JHEP \textbf{11}, 044 (2017)
[arXiv:1706.07808 [hep-ph]].

\bibitem{Kumar:2018kmr}
J.~Kumar, D.~London and R.~Watanabe,
``Combined Explanations of the $b \to s \mu^+ \mu^-$ and $b \to c \tau^- {\bar\nu}$ Anomalies: a General Model Analysis,''
Phys.\ Rev.\ D \textbf{99}, no.1, 015007 (2019)
[arXiv:1806.07403 [hep-ph]].

\bibitem{Angelescu:2021lln}
A.~Angelescu, D.~Be\v{c}irevi\'c, D.~A.~Faroughy, F.~Jaffredo and O.~Sumensari,
``Single leptoquark solutions to the B-physics anomalies,''
Phys.\ Rev.\ D \textbf{104}, no.5, 055017 (2021)
[arXiv:2103.12504 [hep-ph]].

\bibitem{LHCb:2022qnv}
R.~Aaij \textit{et al.} [LHCb],
``Test of lepton universality in $b \rightarrow s \ell^+ \ell^-$ decays,''
Phys.\ Rev.\ Lett.\ \textbf{131}, no.5, 051803 (2023)
[arXiv:2212.09152 [hep-ex]].

\bibitem{LHCb:2022vje}
R.~Aaij \textit{et al.} [LHCb],
``Measurement of lepton universality parameters in $B^+\to K^+\ell^+\ell^-$ and $B^0\to K^{*0}\ell^+\ell^-$ decays,''
Phys.\ Rev.\ D \textbf{108}, no.3, 032002 (2023)
[arXiv:2212.09153 [hep-ex]].

\bibitem{Greljo:2022jac}
A.~Greljo, J.~Salko, A.~Smolkovi\v{c} and P.~Stangl,
``Rare $b$ decays meet high-mass Drell-Yan,''
JHEP \textbf{05}, 087 (2023)
[arXiv:2212.10497 [hep-ph]].

\bibitem{Alguero:2023jeh}
M.~Alguer\'o, A.~Biswas, B.~Capdevila, S.~Descotes-Genon, J.~Matias and M.~Novoa-Brunet,
``To (b)e or not to (b)e: no electrons at LHCb,''
Eur.\ Phys.\ J. C \textbf{83}, no.7, 648 (2023)
[arXiv:2304.07330 [hep-ph]].

\bibitem{Hurth:2023jwr}
T.~Hurth, F.~Mahmoudi and S.~Neshatpour,
``$B$ anomalies in the post $R_{K^{(*)}}$ era,''
Phys.\ Rev.\ D \textbf{108}, no.11, 115037 (2023)
[arXiv:2310.05585 [hep-ph]].

\bibitem{Datta:2013kja}
A.~Datta, M.~Duraisamy and D.~Ghosh,
``Explaining the $B \to K^\ast \mu^+ \mu^-$ data with scalar interactions,''
Phys.\ Rev.\ D \textbf{89}, no.7, 071501 (2014)
[arXiv:1310.1937 [hep-ph]].

\bibitem{Buchalla:1995vs}
G.~Buchalla, A.~J.~Buras and M.~E.~Lautenbacher,
``Weak decays beyond leading logarithms,''
Rev.\ Mod.\ Phys.\ \textbf{68}, 1125-1144 (1996)
[arXiv:hep-ph/9512380 [hep-ph]].

\bibitem{pdg}
S. Navas et al. (Particle Data Group), https://pdg.lbl.gov,
to be published in Phys.\ Rev.\ D \textbf{110}, 030001 (2024)

\bibitem{Gronau:1994rj}
M.~Gronau, O.~F.~Hernandez, D.~London and J.~L.~Rosner,
``Decays of $B$ mesons to two light pseudoscalars,''
Phys.\ Rev.\ D \textbf{50}, 4529-4543 (1994)
[arXiv:hep-ph/9404283 [hep-ph]].

\bibitem{Gronau:1995hn}
M.~Gronau, O.~F.~Hernandez, D.~London and J.~L.~Rosner,
``Electroweak penguins and two-body $B$ decays,''
Phys.\ Rev.\ D \textbf{52}, 6374-6382 (1995)
[arXiv:hep-ph/9504327 [hep-ph]].

\bibitem{Neubert:1998pt}
M.~Neubert and J.~L.~Rosner,
``New bound on $\gamma$ from $B^\pm \to \pi K$ decays,''
Phys.\ Lett.\ B \textbf{441}, 403-409 (1998)
[arXiv:hep-ph/9808493 [hep-ph]].

\bibitem{Neubert:1998jq}
M.~Neubert and J.~L.~Rosner,
``Determination of the weak phase $\gamma$ from rate measurements in $B^\pm \to \pi K, \pi \pi$ decays,''
Phys.\ Rev.\ Lett.\ \textbf{81}, 5076-5079 (1998)
[arXiv:hep-ph/9809311 [hep-ph]].

\bibitem{Gronau:1998fn}
M.~Gronau, D.~Pirjol and T.~M.~Yan,
``Model independent electroweak penguins in $B$ decays to two pseudoscalars,''
Phys.\ Rev.\ D \textbf{60}, 034021 (1999)
[erratum: Phys. Rev. D \textbf{69}, 119901 (2004)]
[arXiv:hep-ph/9810482 [hep-ph]].

\bibitem{Beneke:2001ev}
M.~Beneke, G.~Buchalla, M.~Neubert and C.~T.~Sachrajda,
``QCD factorization in $B \to \pi K, \pi \pi$ decays and extraction of Wolfenstein parameters,''
Nucl.\ Phys.\ B \textbf{606}, 245-321 (2001)
[arXiv:hep-ph/0104110 [hep-ph]].

\bibitem{Bell:2007tv}
G.~Bell,
``NNLO vertex corrections in charmless hadronic $B$ decays: Imaginary part,''
Nucl.\ Phys.\ B \textbf{795}, 1-26 (2008)
[arXiv:0705.3127 [hep-ph]].

\bibitem{Bell:2009nk}
G.~Bell,
``NNLO vertex corrections in charmless hadronic $B$ decays: Real part,''
Nucl.\ Phys.\ B \textbf{822}, 172-200 (2009)
[arXiv:0902.1915 [hep-ph]].

\bibitem{Beneke:2009ek}
M.~Beneke, T.~Huber and X.~Q.~Li,
``NNLO vertex corrections to non-leptonic $B$ decays: Tree amplitudes,''
Nucl.\ Phys.\ B \textbf{832}, 109-151 (2010)
[arXiv:0911.3655 [hep-ph]].

\bibitem{Bell:2015koa}
G.~Bell, M.~Beneke, T.~Huber and X.~Q.~Li,
``Two-loop current-current operator contribution to the non-leptonic QCD penguin amplitude,''
Phys.\ Lett.\ B \textbf{750}, 348-355 (2015)
[arXiv:1507.03700 [hep-ph]].

\bibitem{Kim:2007kx}
For example, see C.~S.~Kim, S.~Oh and Y.~W.~Yoon,
``Analytic resolution of puzzle in $B \to K \pi$ decays,''
Phys.\ Lett.\ B \textbf{665}, 231-236 (2008)
[arXiv:0707.2967 [hep-ph]].

\bibitem{Datta:2004jm}
A.~Datta, M.~Imbeault, D.~London, V.~Pag\'e, N.~Sinha and R.~Sinha,
``Methods for measuring new-physics parameters in $B$ decays,''
Phys.\ Rev.\ D \textbf{71}, 096002 (2005)
[arXiv:hep-ph/0406192 [hep-ph]].

\bibitem{HFLAV:2022esi}
Y.~S.~Amhis \textit{et al.} [HFLAV],
``Averages of b-hadron, c-hadron, and \ensuremath{\tau}-lepton properties as of 2021,''
Phys.\ Rev.\ D \textbf{107}, no.5, 052008 (2023)
[arXiv:2206.07501 [hep-ex]] and online updates at
https://hflav.web.cern.ch.

\bibitem{Baek:2009pa}
S.~Baek, C.~W.~Chiang and D.~London,
``The B ---\ensuremath{>} pi K Puzzle: 2009 Update,''
Phys.\ Lett.\ B \textbf{675}, 59-63 (2009)
[arXiv:0903.3086 [hep-ph]].

\bibitem{Bobeth:2011st}
C.~Bobeth and U.~Haisch,
``New Physics in $\Gamma_{12}^s$: ($\bar{s} b$)$(\bar{\tau} \tau)$ Operators,''
Acta Phys.\ Polon.\ B \textbf{44}, 127-176 (2013)
[arXiv:1109.1826 [hep-ph]].

\bibitem{Aebischer:2018bkb}
J.~Aebischer, J.~Kumar and D.~M.~Straub,
``Wilson: a Python package for the running and matching of Wilson coefficients above and below the electroweak scale,''
Eur.\ Phys.\ J. C \textbf{78}, no.12, 1026 (2018)
[arXiv:1804.05033 [hep-ph]].

\bibitem{Iguro:2024hyk} For the latest update of the $\bctaunu$ anomaly, see
S.~Iguro, T.~Kitahara and R.~Watanabe,
``Global fit to $b \to c\tau\nu$ anomaly 2024 Spring breeze,''
[arXiv:2405.06062 [hep-ph]].

\bibitem{Belle-II:2023esi}
I.~Adachi \textit{et al.} [Belle-II],
``Evidence for B+\textrightarrow{}K+\ensuremath{\nu}\ensuremath{\nu}\textasciimacron{} decays,''
Phys.\ Rev.\ D \textbf{109}, no.11, 112006 (2024)
[arXiv:2311.14647 [hep-ex]].



\bibitem{RBC:2020kdj}
R.~Abbott \textit{et al.} [RBC and UKQCD],
``Direct CP violation and the $\Delta I=1/2$ rule in $K\to\pi\pi$ decay from the standard model,''
Phys.\ Rev.\ D \textbf{102}, no.5, 054509 (2020)
[arXiv:2004.09440 [hep-lat]].

\bibitem{Buras:1999st}
A.~J.~Buras, P.~Gambino and U.~A.~Haisch,
``Electroweak penguin contributions to nonleptonic $\Delta F = 1$ decays at NNLO,''
Nucl.\ Phys.\ B \textbf{570}, 117-154 (2000)
[arXiv:hep-ph/9911250 [hep-ph]].

\bibitem{Buras:2020pjp}
A.~J.~Buras and J.~M.~G\'erard,
``Isospin-breaking in $\varepsilon '/\varepsilon $: impact of $\eta _0$ at the dawn of the 2020s,''
Eur.\ Phys.\ J. C \textbf{80}, no.8, 701 (2020)
[arXiv:2005.08976 [hep-ph]].

\bibitem{NA48:2002tmj}
J.~R.~Batley \textit{et al.} [NA48],
``A Precision measurement of direct CP violation in the decay of neutral kaons into two pions,''
Phys.\ Lett.\ B \textbf{544}, 97-112 (2002)
[arXiv:hep-ex/0208009 [hep-ex]].

\bibitem{KTeV:2002qqy}
A.~Alavi-Harati \textit{et al.} [KTeV],
``Measurements of direct CP violation, CPT symmetry, and other parameters in the neutral kaon system,''
Phys.\ Rev.\ D \textbf{67}, 012005 (2003)
[erratum: Phys.\ Rev.\ D \textbf{70}, 079904 (2004)]
[arXiv:hep-ex/0208007 [hep-ex]].

\bibitem{Worcester:2009qt}
E.~T.~Worcester [KTeV],
``The Final Measurement of Epsilon-prime/Epsilon from KTeV,''
[arXiv:0909.2555 [hep-ex]].

\cite{Aebischer:2021hws}
\bibitem{Aebischer:2021hws}
J.~Aebischer, C.~Bobeth, A.~J.~Buras and J.~Kumar,
``BSM master formula for \ensuremath{\varepsilon}'/\ensuremath{\varepsilon} in the WET basis at NLO in QCD,''
JHEP \textbf{12}, 043 (2021)
[arXiv:2107.12391 [hep-ph]].



\bibitem{Faroughy:2016osc}
D.~A.~Faroughy, A.~Greljo and J.~F.~Kamenik,
``Confronting lepton flavor universality violation in $B$ decays with high-$p_T$ tau lepton searches at LHC,''
Phys.\ Lett.\ B \textbf{764}, 126-134 (2017)
[arXiv:1609.07138 [hep-ph]].

\bibitem{Becirevic:2024pni}
D.~Be\v{c}irevi\'c, S.~Fajfer, N.~Ko\v{s}nik and L.~Pavi\v{c}i\'c,
``$R_{D^{(*)}}$ and survival of the fittest scalar leptoquark,''
[arXiv:2404.16772 [hep-ph]].

\bibitem{Alte:2017pme}
S.~Alte, M.~K\"onig and W.~Shepherd,
``Consistent Searches for SMEFT Effects in Non-Resonant Dijet Events,''
JHEP \textbf{01}, 094 (2018)
[arXiv:1711.07484 [hep-ph]].

\bibitem{Keilmann:2019cbp}
E.~Keilmann and W.~Shepherd,
``Dijets at Tevatron Cannot Constrain SMEFT Four-Quark Operators,''
JHEP \textbf{09}, 086 (2019)
[arXiv:1907.13160 [hep-ph]].

\bibitem{Greljo:2017vvb}
A.~Greljo and D.~Marzocca,
``High-$p_T$ dilepton tails and flavor physics,''
Eur.\ Phys.\ J. C \textbf{77}, no.8, 548 (2017)
[arXiv:1704.09015 [hep-ph]].


\bibitem{Alguero:2022wkd}
M.~Alguer\'o, J.~Matias, B.~Capdevila and A.~Crivellin,
``Disentangling lepton flavor universal and lepton flavor universality violating effects in b\textrightarrow{}s\ensuremath{\ell}+\ensuremath{\ell}- transitions,''
Phys.\ Rev.\ D \textbf{105}, no.11, 113007 (2022)
[arXiv:2205.15212 [hep-ph]].

\end{thebibliography}
\end{document}